\pdfoutput=1
\documentclass[12pt]{article}

\usepackage[T1]{fontenc}
\usepackage[utf8]{inputenc}
\usepackage{textcomp}
\usepackage{lmodern}
\usepackage{amsmath,amssymb,mathtools}
\usepackage{newunicodechar}
\usepackage{geometry}
\usepackage{setspace}
\usepackage{xcolor}
\usepackage{longtable,booktabs,array}
\usepackage{calc}
\usepackage{etoolbox}
\usepackage{graphicx}
\usepackage{microtype}
\usepackage{hyperref}
\usepackage{xurl}
\hypersetup{hidelinks,pdftitle={Design-Indexed Heterogeneity in Meta-Analysis: A Constrained Location-Scale Specification},pdfauthor={Subir Hait}}
\newcommand{\real}[1]{#1}

\makeatletter
\patchcmd\longtable{\par}{\if@noskipsec\mbox{}\fi\par}{}{}
\makeatother
\setkeys{Gin}{keepaspectratio}

\newunicodechar{γ}{\ensuremath{\gamma}}
\newunicodechar{τ}{\ensuremath{\tau}}
\newunicodechar{μ}{\ensuremath{\mu}}
\newunicodechar{β}{\ensuremath{\beta}}
\newunicodechar{α}{\ensuremath{\alpha}}
\newunicodechar{δ}{\ensuremath{\delta}}
\newunicodechar{ε}{\ensuremath{\varepsilon}}
\newunicodechar{η}{\ensuremath{\eta}}
\newunicodechar{σ}{\ensuremath{\sigma}}
\newunicodechar{ψ}{\ensuremath{\psi}}
\newunicodechar{Σ}{\ensuremath{\Sigma}}
\newunicodechar{ᵢ}{\ensuremath{_{i}}}
\newunicodechar{ⱼ}{\ensuremath{_{j}}}
\newunicodechar{₀}{\ensuremath{_{0}}}
\newunicodechar{₁}{\ensuremath{_{1}}}
\newunicodechar{ₖ}{\ensuremath{_{k}}}
\newunicodechar{²}{\ensuremath{^{2}}}
\newunicodechar{¹}{\ensuremath{^{1}}}
\newunicodechar{⁻}{\ensuremath{^{-}}}
\newunicodechar{′}{\ensuremath{^{\prime}}}
\newunicodechar{×}{\ensuremath{\times}}
\newunicodechar{∈}{\ensuremath{\in}}
\newunicodechar{∑}{\ensuremath{\sum}}
\newunicodechar{−}{\ensuremath{-}}
\newunicodechar{∗}{\ensuremath{\ast}}
\newunicodechar{∞}{\ensuremath{\infty}}
\newunicodechar{≈}{\ensuremath{\approx}}
\newunicodechar{≤}{\ensuremath{\leq}}
\newunicodechar{≥}{\ensuremath{\geq}}
\newunicodechar{†}{\ensuremath{\dagger}}
\newunicodechar{□}{\ensuremath{\square}}

\begin{document}
\begin{singlespace}
\begin{center}
{\Large\bfseries Design-Indexed Heterogeneity in Meta-Analysis:\\[0.2em]
A Constrained Location-Scale Specification\par}
\vspace{0.75em}
{\large Subir Hait\par}
\vspace{0.25em}
Counseling, Educational Psychology and Special Education\\
Michigan State University\\[0.25em]
ORCID: 0009-0004-9871-9677\\
Correspondence: \href{mailto:haitsubi@msu.edu}{haitsubi@msu.edu}
\end{center}
\end{singlespace}

\clearpage
\begin{samepage}
\noindent{\large\bfseries Abstract}\par
\vspace{0.3em}
An analyst should not assume that a design score improves a
meta-analysis simply because it is labeled as robustness. This paper
treats declining heterogeneity along a prespecified, outcome-separated
design score as a falsifiable directional hypothesis. In an 18-study
illustration, the constrained model lands at γ = 0 and becomes
numerically identical to conventional random effects, while the
unrestricted scale fit points in the opposite direction. Simulations
that deliberately violate the restriction show the same graceful
fallback under wrong-direction and U-shaped heterogeneity; a stepwise
decreasing variance function retains an efficiency gain despite
functional-form misspecification. DR-Meta formalizes the hypothesis as
τ²(DR) = τ₀² exp(\ensuremath{-}γDR), γ ≥ 0, within an established location-scale
framework. Exact nesting, conditional weight monotonicity, a
fixed-effect upper bound, pseudo-true targets under mean
misspecification, and score-rescaling behavior are derived with full
proofs. Efficiency gains are negligible around an empirically anchored γ
near 1 and become material only under stronger gradients; the largest
RMSE reduction, about 9.6\%, occurs in a deliberately extreme condition.
Plug-in intervals under-cover relative to random effects, and modified
Knapp-Hartung adjustment improves coverage only modestly. Widening the γ
optimization bound changes γ estimates much more than pooled-estimate
RMSE. The practical contribution is therefore diagnostic rather than
automatic efficiency: a nested directional restriction can be supported,
contradicted, or collapse transparently to random effects, provided
constrained, unrestricted, location, and sensitivity analyses are
reported together.

\vspace{0.45em}
\end{samepage}

\clearpage
\section{\texorpdfstring{\textbf{1.
Introduction}}{1. Introduction}}\label{introduction}

Conventional random-effects meta-analysis allows true effects to vary
but ordinarily treats the residual between-study variance as constant
after any location moderators have been included.\textsuperscript{1-3}

That exchangeability assumption is convenient, yet it can be
scientifically awkward when a synthesis combines studies that differ
materially in randomization integrity, covariate balance, overlap,
attrition control, measurement quality, or other design features that
may be associated with the stability of estimated effects.

Design labels nevertheless do not determine the magnitude or direction
of bias. Generic quality scores can be subjective, multidimensional, and
only weakly related to the feature they are intended to summarize. The
purpose of DR-Meta is therefore not to turn a design ranking into a
causal correction, but to ask a narrower scale question while preserving
explicit guardrails against that interpretation.\textsuperscript{4-6}

General meta-analytic location-scale models already allow covariates to
predict both the conditional mean and residual heterogeneity, and recent
work has evaluated their estimation and finite-sample
behavior.\textsuperscript{7,8}

DR-Meta is not proposed as a new general location-scale model. It is a
constrained specialization for a narrower scientific question: does
residual heterogeneity stay the same or decrease along a prespecified
design-robustness dimension?

This distinction matters because quality weighting and bias adjustment
answer different questions. Quality-effects methods alter weights using
external study information, whereas bias models attempt to represent
systematic location shifts. Composite scores may themselves introduce
distortion if their components are poorly chosen or if realized outcomes
influence the score.\textsuperscript{4-6,9}

The proposed model writes residual heterogeneity as τ²(DR) = τ₀²
exp(\ensuremath{-}γDR) and constrains γ ≥ 0. The restriction says only that,
conditional on the declared model, unexplained heterogeneity is not
allowed to increase with design robustness. It does not say that a
higher DR score removes bias, identifies a causal effect of ``better
design,'' or guarantees that high-DR studies should receive the largest
total weight.

The analysis deliberately subjects that restriction to conditions in
which it is wrong. It also examines interval coverage, includes an
empirically anchored scale-gradient condition, widens the optimization
bound when γ is weakly identified, and includes an empirical
illustration in which the constrained model reaches γ = 0 rather than
manufacturing a favorable gradient.

The paper makes five contributions. First, it gives the constrained
model a precise estimand and shows how it nests standard random-effects
analysis. Second, it derives formal weight-ordering and boundedness
properties. Third, it develops observed-support scale summaries that do
not masquerade as an explained-heterogeneity R². Fourth, it evaluates
the model under correct specification, score degradation, location
misspecification, scale misspecification, and boundary stress. Fifth, it
provides a practical workflow in which constrained, unrestricted,
location, and sensitivity fits are reported together.

\section{\texorpdfstring{\textbf{2. Relationship to Existing
Meta-Analytic
Models}}{2. Relationship to Existing Meta-Analytic Models}}\label{relationship-to-existing-meta-analytic-models}

\subsection{\texorpdfstring{\textbf{2.1 Conventional random-effects
meta-analysis}}{2.1 Conventional random-effects meta-analysis}}\label{conventional-random-effects-meta-analysis}

Let yᵢ denote an approximately unbiased study-level effect estimate with
known or estimated sampling variance vᵢ \textgreater{} 0. The
conventional intercept-only random-effects model is

\begin{equation*}
y_{i} = \mu + u_{i} + e_{i},\quad\quad u_{i}\mathcal{\sim N}\left( 0,\tau^{2} \right),\quad\quad e_{i}\mathcal{\sim N}\left( 0,v_{i} \right).\tag{1}
\end{equation*}

Conditional on τ², the pooled estimator is an inverse-total-variance
average with weights 1/(vᵢ + τ²). A meta-regression replaces μ by xᵢ′β,
but residual between-study variance remains constant unless a scale
model is introduced.

\subsection{\texorpdfstring{\textbf{2.2 Quality weighting, design
information, and bias
adjustment}}{2.2 Quality weighting, design information, and bias adjustment}}\label{quality-weighting-design-information-and-bias-adjustment}

Quality-effects procedures explicitly modify study weights according to
external quality information.\textsuperscript{9}

That approach is conceptually distinct from DR-Meta. A scalar quality
score can combine design dimensions with different relations to bias and
variance, and the hazards of using composite scores as if they were
direct bias measurements are well documented.\textsuperscript{4-6}

DR-Meta therefore requires a declared design-robustness score and uses
it only as a scale moderator in the primary model. If a systematic mean
shift with DR is plausible, that shift belongs in the location model or
in a separate bias analysis.

Quality-effects weighting can be useful as an explicitly alternative
weighting scheme, but its estimand and likelihood differ from a
variance-function model.6 Multiplying inverse-variance weights by a
quality score does not, by itself, identify a bias process or establish
that the resulting weights correspond to the conditional variance of
study effects. DR-Meta instead makes the scale relation explicit and
falsifiable through the constrained-versus-unrestricted comparison.

\subsection{\texorpdfstring{\textbf{2.3 Location-scale
meta-analysis}}{2.3 Location-scale meta-analysis}}\label{location-scale-meta-analysis}

In a meta-analytic location-scale model, study-level covariates can
predict both the conditional mean and the log of residual between-study
heterogeneity. A general specification is

\begin{equation*}
y_{i} \mid \mathbf{x}_{i},\mathbf{z}_{i}\mathcal{\sim N}\left( \mathbf{x}_{i}^{\top}\mathbf{\beta},\mspace{6mu} v_{i} + \exp\left( \mathbf{z}_{i}^{\top}\mathbf{\alpha} \right) \right).\tag{2}
\end{equation*}

DR-Meta is obtained by setting zᵢ = (1, DRᵢ), writing α₀ = log(τ₀²), and
constraining the coefficient on DRᵢ to \ensuremath{-}γ with γ ≥ 0. The general
location-scale model is therefore the statistical parent model; the
contribution here is the directional design-indexed restriction, its
diagnostics, and its reporting discipline.\textsuperscript{7,8}

The unrestricted signed counterpart does not require a separate software
ecosystem. In metafor, a location-scale model can be fitted with
rma(..., scale = \textasciitilde{} DR) under the default log link, so
that log residual heterogeneity is modeled as α₀ + α₁DR. DR-Meta uses
the same parent model with α₁ = \ensuremath{-}γ and the one-sided restriction γ ≥ 0.
The drmeta implementation is therefore not presented as a replacement
for metafor; its software role is to package the directional constraint,
exact γ = 0 nesting, observed-support scale summaries, and the
constrained-versus-unrestricted diagnostic workflow.\textsuperscript{7}

\subsection{\texorpdfstring{\textbf{2.4 What the design-robustness index
represents}}{2.4 What the design-robustness index represents}}\label{what-the-design-robustness-index-represents}

DRᵢ is not a universal ranking of study designs. It is a study-level
summary constructed for a specific synthesis, outcome, and
identification problem. Potential primary inputs include randomization
integrity, assignment procedures, covariate balance, overlap, attrition,
measurement quality, prespecification, and other design features that
can be coded without using the realized meta-analytic effect.

\textbf{Primary-score separation condition.} For confirmatory use, DRᵢ
should be constructed without using yᵢ, its standard error, p value,
confidence interval, or any diagnostic that is itself a function of
those realized outcome quantities. Outcome-dependent sensitivity
measures may be scientifically useful, but a score containing them
should be labeled exploratory rather than treated as the primary scale
moderator.

A study-level association between DR and heterogeneity remains
observational across studies. DR-Meta asks whether residual dispersion
is patterned along the declared design dimension; it does not imply that
intervening to change a study's DR score would causally change
heterogeneity.

\section{\texorpdfstring{\textbf{3. Model, Estimands, and
Interpretation}}{3. Model, Estimands, and Interpretation}}\label{model-estimands-and-interpretation}

\subsection{\texorpdfstring{\textbf{3.1 Scale-only
DR-Meta}}{3.1 Scale-only DR-Meta}}\label{scale-only-dr-meta}

For k independent studies, the scale-only DR-Meta model is

\begin{equation*}
y_{i} \mid DR_{i}\mathcal{\sim N}\left( \mu,\sigma_{i}^{2} \right),\quad\quad\sigma_{i}^{2} = v_{i} + \tau_{0}^{2}\exp\left( - \gamma DR_{i} \right),\tag{3}
\end{equation*}

\begin{equation*}
\tau_{0}^{2} \geq 0,\quad\quad\gamma \geq 0,\quad\quad DR_{i} \in \lbrack 0,1\rbrack.\tag{4}
\end{equation*}

Here τ₀² is the residual between-study variance at DR = 0, and γ
controls its proportional decay with increasing DR. At score value d,
τ²(d) = τ₀² exp(\ensuremath{-}γd). Conditional on the estimated scale parameters, the
pooled estimator is

\begin{equation*}
{\widehat{\mu}}_{DR} = \frac{\sum_{i}^{}w_{i}y_{i}}{\sum_{i}^{}w_{i}},\quad\quad w_{i} = \frac{1}{v_{i} + \tau_{0}^{2}\exp\left( - \gamma DR_{i} \right)}.\tag{5}
\end{equation*}

When the common-mean model is correctly specified, μ is the mean shared
by the conditional distributions and the scale function determines how
studies contribute precision. When the mean varies with omitted study
characteristics, \ensuremath{\hat{\mu}}DR is instead a model-dependent weighted average.

\subsection{\texorpdfstring{\textbf{3.2 Joint design-indexed
location-scale
model}}{3.2 Joint design-indexed location-scale model}}\label{joint-design-indexed-location-scale-model}

When a design-related mean shift is scientifically plausible, the
corresponding joint location-scale model is

\begin{equation*}
y_{i} \mid \mathbf{x}_{i},DR_{i}\mathcal{\sim N}\left( \mathbf{x}_{i}^{\top}\mathbf{\beta},\mspace{6mu} v_{i} + \tau_{0}^{2}\exp\left( - \gamma DR_{i} \right) \right),\quad\quad\gamma \geq 0.\tag{6}
\end{equation*}

A simple location specification uses xᵢ = (1, 1 - DRᵢ). Under that
parameterization, β₀ is the fitted mean at DR = 1 and β₁ is the fitted
difference between DR = 0 and DR = 1 under a linear mean relation. The
high-DR intercept is an extrapolated statistical target whenever few
studies lie near DR = 1 and should not automatically be interpreted as
the causal effect of a hypothetical perfect study.

\subsection{\texorpdfstring{\textbf{3.3 Pseudo-true target under a
misspecified constant
mean}}{3.3 Pseudo-true target under a misspecified constant mean}}\label{pseudo-true-target-under-a-misspecified-constant-mean}

Suppose the true conditional mean is m(DRᵢ), but the fitted scale-only
model imposes a constant mean. For fixed variance parameters, the
population objective yields the pseudo-true target

\begin{equation*}
\mu^{\star} = \frac{E\left\lbrack w(DR,v)m(DR) \right\rbrack}{E\left\lbrack w(DR,v) \right\rbrack}.\tag{7}
\end{equation*}

This expression shows both the possible value and the limitation of
design-indexed weighting. If lower-DR studies tend to have mean
distortion in one direction and receive less weight, μ* can move toward
a high-DR target. There is no general unbiasedness result: the change
depends jointly on the mean function, sampling variances, residual
heterogeneity, and the DR distribution.

\section{\texorpdfstring{\textbf{4. Formal
Properties}}{4. Formal Properties}}\label{formal-properties}

\subsection{\texorpdfstring{\textbf{4.1 Nesting and
identification}}{4.1 Nesting and identification}}\label{nesting-and-identification}

Proposition 1 (nesting). If γ = 0, Equation 3 is exactly the
conventional random-effects model with τ² = τ₀². If τ₀² = 0, the model
reduces to the common-effect model, or to fixed-effects meta-regression
when xᵢ contains moderators. If DRᵢ = c for every study, \ensuremath{\hat{\mu}}DR equals the
conventional random-effects pooled estimate with τ²(c) = τ₀² exp(\ensuremath{-}γc).
In this constant-DR case, τ₀² and γ are not separately identifiable.

The constant-DR result identifies the practical support requirement for
γ: nondegenerate DR variation, enough residual heterogeneity to reveal
scale differences, and enough studies distributed across the observed
score range. The pooled location can remain stable while γ is weakly
identified.

\subsection{\texorpdfstring{\textbf{4.2 Monotone weight
ordering}}{4.2 Monotone weight ordering}}\label{monotone-weight-ordering}

Proposition 2 (conditional monotonicity). For equal sampling variances
vᵢ = vⱼ and DRᵢ \textgreater{} DRⱼ, DR-Meta weights satisfy wᵢ ≥ wⱼ
whenever γ ≥ 0. The inequality is strict when γ \textgreater{} 0 and τ₀²
\textgreater{} 0. Thus, higher DR produces greater weight conditional on
equal sampling variance only when there is positive residual
heterogeneity and a nonzero scale gradient. A lower-DR study with much
smaller sampling variance can still receive more total weight.

\subsection{\texorpdfstring{\textbf{4.3
Boundedness}}{4.3 Boundedness}}\label{boundedness}

Proposition 3 (fixed-effect upper bound). Because τ₀² exp(\ensuremath{-}γDRᵢ) ≥ 0,
every DR-Meta weight satisfies 0 \textless{} wᵢ ≤ 1/vᵢ. The
variance-function model therefore cannot assign a study more raw
inverse-variance weight than it would receive under common-effect
weighting.

\subsection{\texorpdfstring{\textbf{4.4 Scale attenuation over observed
support}}{4.4 Scale attenuation over observed support}}\label{scale-attenuation-over-observed-support}

The scale model directly specifies conditional residual between-study
variance as a function of DR. Interpretation is therefore based on
contrasts in the fitted variance function rather than on an additive
``design-explained'' heterogeneity decomposition. For lower and upper
score values dL and dH, define

\begin{equation*}
R_{\tau}\left( d_{L},d_{H} \right) = \frac{\tau^{2}\left( d_{H} \right)}{\tau^{2}\left( d_{L} \right)} = \exp\left\lbrack - \gamma\left( d_{H} - d_{L} \right) \right\rbrack,\quad\quad A_{\tau}\left( d_{L},d_{H} \right) = 1 - R_{\tau}\left( d_{L},d_{H} \right).\tag{8}
\end{equation*}

For τ₀² \textgreater{} 0, Rτ is the fitted residual-heterogeneity
variance ratio and Aτ the corresponding proportional attenuation. Under
γ ≥ 0 and dH ≥ dL, Rτ lies in (0,1{]} and Aτ in {[}0,1). These
quantities describe the fitted scale function; they are not causal R²
measures or fractions of heterogeneity ``explained'' by design quality.

Because γ changes under rescaling of DR, substantive reporting should
pair γ with the observed DR range and with Rτ or Aτ evaluated over a
prespecified contrast within observed support.

\subsection{\texorpdfstring{\textbf{4.5 Interior asymptotics and the
boundary}}{4.5 Interior asymptotics and the boundary}}\label{interior-asymptotics-and-the-boundary}

At an identifiable interior point with τ₀² \textgreater{} 0 and γ
\textgreater{} 0, standard smooth-likelihood arguments apply under usual
regularity conditions. At γ = 0 the constrained parameter lies on the
boundary, so ordinary two-sided Wald and interior chi-square
likelihood-ratio references are not automatically
justified.\textsuperscript{10}

Appendix A provides full mathematical proofs and derivations rather than
proof sketches.

\section{\texorpdfstring{\textbf{5. Estimation, Inference, and
Diagnostics}}{5. Estimation, Inference, and Diagnostics}}\label{estimation-inference-and-diagnostics}

\subsection{\texorpdfstring{\textbf{5.1 Profiled likelihood
estimation}}{5.1 Profiled likelihood estimation}}\label{profiled-likelihood-estimation}

Let X denote the k by p location-model matrix and Σ(ψ) = diag{[}vᵢ + τ₀²
exp(\ensuremath{-}γDRᵢ){]}, with ψ = (τ₀², γ). Conditional on ψ, generalized least
squares gives

\begin{equation*}
\widehat{\mathbf{\beta}}\left( \mathbf{\psi} \right) = \left\lbrack \mathbf{X}^{\top}\mathbf{\Sigma}\left( \mathbf{\psi} \right)^{- 1}\mathbf{X} \right\rbrack^{- 1}\mathbf{X}^{\top}\mathbf{\Sigma}\left( \mathbf{\psi} \right)^{- 1}\mathbf{y}.\tag{9}
\end{equation*}

Estimation profiles β out of the ML or REML criterion and optimizes the
scale parameters. Numerically, log(τ₀²) is used for positivity while γ
is optimized directly on a finite interval that contains 0. This allows
the exact γ = 0 boundary to be reached.

\subsection{\texorpdfstring{\textbf{5.2 Boundary-aware inference for
γ}}{5.2 Boundary-aware inference for γ}}\label{boundary-aware-inference-for-ux3b3}

Because γ = 0 is a boundary point, a parametric-bootstrap comparison of
γ = 0 with γ ≥ 0 is a practical inferential option. The unrestricted
scale fit, which lets the DR coefficient on log heterogeneity take
either sign, is a complementary directional
diagnostic.\textsuperscript{10}

This article does not claim finite-sample size or power calibration for
the parametric-bootstrap γ test. The inferential recommendation is
retained as a boundary-aware option, while the empirical evidence in
this paper focuses on estimation, constrained-versus-unrestricted
directional diagnosis, coverage, misspecification, and bound
sensitivity.

If the unrestricted fit estimates increasing heterogeneity with DR, the
substantive constraint is contradicted; the constrained estimator should
then be expected to accumulate at γ = 0 rather than force a declining
variance relation.

\subsection{\texorpdfstring{\textbf{5.3 Location uncertainty and
modified Knapp-Hartung
intervals}}{5.3 Location uncertainty and modified Knapp-Hartung intervals}}\label{location-uncertainty-and-modified-knapp-hartung-intervals}

Plug-in standard errors based on (X′\ensuremath{\hat{\Sigma}}\ensuremath{^{-1}}X)\ensuremath{^{-1}} condition on the estimated
scale parameters and can undercover when the number of studies is modest
or γ is weakly identified. Knapp-Hartung-type adjustments are standard
small-sample tools in meta-regression.\textsuperscript{11}

The adjustment is deliberately treated as a comparator rather than a
complete solution. Knapp-Hartung-type procedures address small-sample
uncertainty in meta-regression coefficients,12 but a plug-in use of the
fitted location-scale covariance still does not integrate over
uncertainty in both τ₀² and γ. Profile-likelihood or bootstrap intervals
remain preferable when finite-sample coverage of the location parameter
is the primary inferential target.

For the intercept-only comparisons below, the modified Knapp-Hartung
(mKH) interval multiplies the GLS location variance by max(1, q), where
q = Σwᵢ(yᵢ - \ensuremath{\hat{\mu}})²/(k - 1), and uses a t reference with k - 1 degrees of
freedom. The max(1, q) modification prevents the adjustment from
producing an interval narrower than the plug-in t interval. Because this
adjustment does not fully propagate uncertainty in τ₀² and γ, it is
evaluated as a pragmatic comparator rather than presented as an exact
solution.

\subsection{\texorpdfstring{\textbf{5.4 Recommended
diagnostics}}{5.4 Recommended diagnostics}}\label{recommended-diagnostics}

\begin{itemize}
\item
  Display the empirical distribution, observed range, and number of
  distinct DR values.
\item
  Plot fitted τ²(DR) over observed support rather than extrapolating
  automatically to {[}0,1{]}.
\item
  Report γ with an observed-support variance ratio or attenuation
  contrast, convergence information, and boundary status.
\item
  Compare constrained DR-Meta with conventional random effects and an
  unrestricted scale model.
\item
  Fit an explicit location model whenever design-linked mean differences
  are plausible.
\item
  Repeat analyses under defensible alternative DR constructions and with
  influential studies removed.
\item
  If estimates accumulate at the upper optimization bound, widen that
  bound and verify whether the pooled location materially changes.
\end{itemize}

Figure 1 visualizes the prespecified exponential scale family: γ = 1
represents the empirically anchored region, whereas γ = 4 serves as a
strong-gradient stress condition.

Figure 1. Constrained exponential variance functions for τ₀² = 0.10. γ =
1 is close to the external calibration discussed in Section 6.2; γ = 4
is a deliberately strong-gradient condition rather than a representative
default.

\begin{center}
\includegraphics[width=0.96\linewidth,keepaspectratio]{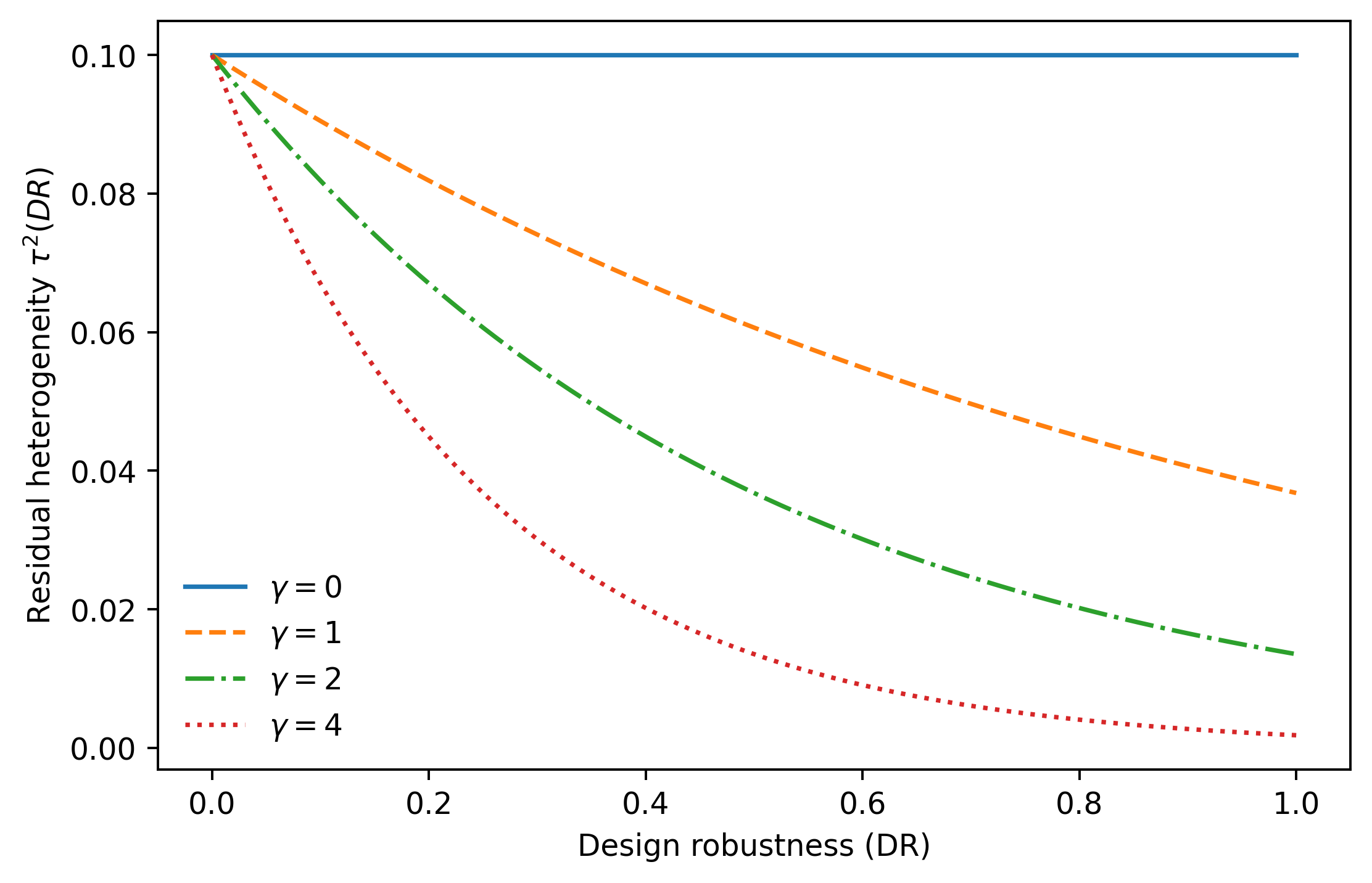}
\end{center}

\section{\texorpdfstring{\textbf{6. Simulation
Study}}{6. Simulation Study}}\label{simulation-study}

\subsection{\texorpdfstring{\textbf{6.1 General
design}}{6.1 General design}}\label{general-design}

Simulation design follows recommendations to state estimands,
data-generating mechanisms, Monte Carlo size, and performance criteria
explicitly.\textsuperscript{12}

In the main factorial experiment, the target common mean is μ = 0.30.
For each replicate, DRᵢ \textasciitilde{} Beta(2,2), sampling variances
vᵢ \textasciitilde{} Uniform(0.005, 0.030), and yᵢ \textasciitilde{}
Normal{[}μ, vᵢ + τ₀² exp(\ensuremath{-}γDRᵢ){]}. We cross k ∈ \{20, 50\}, τ₀² ∈
\{0.02, 0.10\}, and γ ∈ \{0, 1, 2, 4\}. Each main condition uses 1,000
Monte Carlo replications. The competing estimators are conventional
random effects (RE), constrained DR-Meta (DR), and an unrestricted
location-scale fit (LS).

The primary performance measures are bias, RMSE, empirical 95\%
coverage, convergence, γ recovery, and the ratio RMSE(DR)/RMSE(RE).
Values below 1 favor DR-Meta for point estimation. Both plug-in t
intervals and mKH intervals are evaluated.

\subsection{\texorpdfstring{\textbf{6.2 External calibration of the
scale-gradient
grid}}{6.2 External calibration of the scale-gradient grid}}\label{external-calibration-of-the-scale-gradient-grid}

The strong-gradient condition γ = 4 implies exp(-4) = 0.018
residual-variance retention over a full DR-unit contrast, or
approximately 98.2\% attenuation. This condition is intentionally
extreme. Kuper and colleagues reported residual heterogeneity of τ² =
0.712 for high-risk-of-bias studies and τ² = 0.243 for low-risk-of-bias
studies. Mapping high risk to dL = 0 and low risk to dH = 1 under τ²(d)
= τ₀² exp(\ensuremath{-}γd) gives γ = log(0.712/0.243)/(1 - 0) = 1.075, approximately
1.08.\textsuperscript{13}

Accordingly, γ = 1 serves as an empirically anchored moderate condition,
γ = 2 is treated as strong, and γ = 4 serves only as a stress condition.
The largest γ = 4 improvement is therefore interpreted as an
extreme-case result rather than as typical performance.

\subsection{\texorpdfstring{\textbf{6.3 Score-quality and
location-misspecification
experiments}}{6.3 Score-quality and location-misspecification experiments}}\label{score-quality-and-location-misspecification-experiments}

The score-quality experiment uses k = 50, τ₀² = 0.10, and true γ = 4.
Each generated dataset is analyzed with the exact latent DR value, DR
plus independent N(0, 0.15²) error truncated to {[}0,1{]}, and a
three-level coarsening taking values 0, 0.5, and 1. Common random
numbers keep the effect estimates and sampling variances identical
across score conditions, so the conventional RE comparator is the same
by construction. The location-misspecification experiment uses k = 50,
τ₀²(DR) = 0.10 exp(-3DR), and E(yᵢ \textbar{} DRᵢ) = μ + δ(1 - DRᵢ),
with δ ∈ \{0, 0.10, 0.20\}. It compares RE, scale-only DR-Meta,
mean-only random-effects meta-regression (MR), and the joint constrained
location-scale model (JLS). MR and JLS use 1 - DR as the location
moderator, so their intercept is the fitted mean at DR = 1.

\subsection{\texorpdfstring{\textbf{6.4 Scale-model misspecification
experiment}}{6.4 Scale-model misspecification experiment}}\label{scale-model-misspecification-experiment}

This experiment deliberately violates the fitted exponential-decreasing
model. It uses k = 50, μ = 0.30, DRᵢ \textasciitilde{} Beta(2,2), vᵢ
\textasciitilde{} Uniform(0.005, 0.030), and 1,000 replications per
data-generating mechanism. Three residual-heterogeneity functions are
examined:

\begin{itemize}
\item
  Wrong direction: τ²(d) = 0.10 exp{[}1.5(d - 0.5){]}, so heterogeneity
  increases with DR and directly contradicts γ ≥ 0.
\item
  U-shaped: τ²(d) = 0.03 + 0.28(d - 0.5)², so neither a monotone
  increase nor decrease is correct.
\item
  Step decrease: τ²(d) = 0.10 for d \textless{} 0.5 and 0.025 for d ≥
  0.5, so the direction is correct but the exponential functional form
  is wrong.
\end{itemize}

RE, constrained DR-Meta, and unrestricted LS are compared using the same
point-estimation and interval metrics.

\subsection{\texorpdfstring{\textbf{6.5 Optimization-bound
sensitivity}}{6.5 Optimization-bound sensitivity}}\label{optimization-bound-sensitivity}

Because the main γ = 4 fits showed nontrivial concentration at the
artificial upper bound γ = 8, a separate paired Monte Carlo sensitivity
experiment refits γ = 4 conditions with γmax = 8 and γmax = 20. Within
each sensitivity replicate, the generated dataset is identical across
the two bounds; the sensitivity datasets are independent of the main
factorial datasets in Tables 1 and 2. The purpose is to isolate
sensitivity to scale-gradient truncation rather than to reproduce the
main-run marginal summaries.

\subsection{\texorpdfstring{\textbf{6.6 Implementation and Monte Carlo
reporting}}{6.6 Implementation and Monte Carlo reporting}}\label{implementation-and-monte-carlo-reporting}

Computational provenance is reported by experiment. The score-quality
and design-linked mean-misspecification experiments in Tables 5 and 6
were fitted in R 4.6.0 using drmeta version 0.2.2 with master seed
20260806. For the expanded main factorial experiment (Tables 1 and 2 and
Table B1), the scale-model misspecification experiment (Table 3), the
paired optimization-bound experiment (Table 4), the optimizer-start
diagnostics, and the empirical reference fits, the manuscript archive
retains the exact fit-level outputs from an independently implemented
profiled REML reference run. Supplied R scripts recompute the reported
summaries and figures from those archived outputs. The CRAN release and
versioned GitHub release v0.2.2 identify the public R implementation of
the model. All analyses use the declared location-scale likelihood, and
the constrained fits permit the exact lower boundary γ = 0.\textsuperscript{14,15}

For the archived reference optimization runs summarized in Tables 1-4,
the interior scale-gradient initialization was γ = 1. Because a weakly
identified scale likelihood can satisfy a numerical stopping rule
without moving materially from its start, the archived replication
output was audited for exact nonmovement at γ = 1 in addition to the
lower and upper boundaries. This diagnostic is reported explicitly below
rather than interpreted as parameter recovery.

Performance summaries report bias, RMSE, empirical standard deviation,
model-based standard errors, coverage, convergence, and boundary
behavior. Plug-in intervals use t reference values; mKH intervals are
reported as an additional small-sample comparator. Monte Carlo
conclusions are based on converged finite fits, with convergence and
boundary frequencies reported separately rather than silently discarding
difficult conditions.

The manuscript-specific reproducibility bundle separates numerical
provenance rather than attributing all results to one engine. It
contains the exact archived fit-level outputs for the main factorial,
scale-misspecification, and bound-sensitivity experiments; the R/drmeta
0.2.2 package scripts and recovered package-run summaries for the
score-quality and mean-misspecification experiments; full-precision
empirical inputs; Table 7 and leave-one-out outputs; R scripts that
recompute the reported summaries and regenerate Figures 1-8; a run
manifest; a detailed provenance map; and SHA-256 checksums.
A provenance file maps every table and figure to its source artifact.
The manuscript-specific reproducibility bundle is included with this arXiv submission as ancillary material.

\section{\texorpdfstring{\textbf{7. Simulation
Results}}{7. Simulation Results}}\label{simulation-results}

\subsection{\texorpdfstring{\textbf{7.1 Point-estimation efficiency is
conditional, not
automatic}}{7.1 Point-estimation efficiency is conditional, not automatic}}\label{point-estimation-efficiency-is-conditional-not-automatic}

Table 1 summarizes the high-heterogeneity conditions (τ₀² = 0.10), where
a scale relation has the greatest opportunity to matter. The moderate γ
= 1 condition produces little or no efficiency advantage: DR-Meta is
1.7\% worse than RE at k = 20 and about 0.5\% better at k = 50. At γ =
2, the RMSE reduction is about 1.8\% for k = 20 and 3.2\% for k = 50.
The largest reduction, approximately 9.6\%, occurs only for k = 50, τ₀²
= 0.10, γ = 4. Accordingly, the approximately 9.6\% reduction belongs to
an extreme cell, not to the empirically anchored region.

Absolute bias is small throughout the correctly specified
high-heterogeneity grid. The constrained model pays a modest price when
the gradient is absent: RMSE/RE is 1.030 at k = 20 and 1.008 at k = 50
for γ = 0. At the empirically anchored γ = 1 condition, the ratio is
1.017 at k = 20 and 0.995 at k = 50. At γ = 2, DR-Meta improves to 0.982
and 0.968, and at γ = 4 to 0.959 and 0.904. The unrestricted LS fit is
similar when a positive gradient is present but generally pays a larger
price at γ = 0 because it estimates a signed scale slope rather than
enforcing the prespecified direction.

These comparisons separate the statistical question from the headline
number. The method is not an automatic precision improvement over random
effects. Its efficiency advantage appears when the residual scale
relation is both sufficiently strong and sufficiently informed by the
number of studies and the amount of heterogeneity.

\clearpage
Table 1. Main factorial results under substantial heterogeneity (τ₀² =
0.10)

\textbf{Panel A. Point-estimation performance}

\begin{longtable}[]{@{}
  >{\raggedright\arraybackslash}p{(\columnwidth - 8\tabcolsep) * \real{0.1262}}
  >{\raggedright\arraybackslash}p{(\columnwidth - 8\tabcolsep) * \real{0.1650}}
  >{\raggedright\arraybackslash}p{(\columnwidth - 8\tabcolsep) * \real{0.2330}}
  >{\raggedright\arraybackslash}p{(\columnwidth - 8\tabcolsep) * \real{0.2330}}
  >{\raggedright\arraybackslash}p{(\columnwidth - 8\tabcolsep) * \real{0.2427}}@{}}
\toprule\noalign{}
\begin{minipage}[b]{\linewidth}\raggedright
\textbf{γ}
\end{minipage} & \begin{minipage}[b]{\linewidth}\raggedright
\textbf{Fit}
\end{minipage} & \begin{minipage}[b]{\linewidth}\raggedright
\textbf{Bias}
\end{minipage} & \begin{minipage}[b]{\linewidth}\raggedright
\textbf{RMSE}
\end{minipage} & \begin{minipage}[b]{\linewidth}\raggedright
\textbf{RMSE/RE}
\end{minipage} \\
\midrule\noalign{}
\endhead
\bottomrule\noalign{}
\endlastfoot
\multicolumn{5}{@{}>{\raggedright\arraybackslash}p{(\columnwidth - 8\tabcolsep) * \real{1.0000} + 8\tabcolsep}@{}}{%
\textbf{k = 20}} \\
0 & RE & -0.0040 & 0.07528 & 1.000 \\
0 & DR & -0.0036 & 0.07751 & 1.030 \\
0 & LS & -0.0037 & 0.07823 & 1.039 \\
1 & RE & 0.0002 & 0.06230 & 1.000 \\
1 & DR & 0.0008 & 0.06334 & 1.017 \\
1 & LS & 0.0013 & 0.06501 & 1.043 \\
2 & RE & -0.0017 & 0.05504 & 1.000 \\
2 & DR & -0.0022 & 0.05404 & 0.982 \\
2 & LS & -0.0025 & 0.05462 & 0.992 \\
4 & RE & 0.0001 & 0.04201 & 1.000 \\
4 & DR & -0.0003 & 0.04027 & 0.959 \\
4 & LS & -0.0004 & 0.04079 & 0.971 \\
\multicolumn{5}{@{}>{\raggedright\arraybackslash}p{(\columnwidth - 8\tabcolsep) * \real{1.0000} + 8\tabcolsep}@{}}{%
\textbf{k = 50}} \\
0 & RE & -0.0010 & 0.04937 & 1.000 \\
0 & DR & -0.0010 & 0.04979 & 1.008 \\
0 & LS & -0.0013 & 0.05032 & 1.019 \\
1 & RE & -0.0011 & 0.03958 & 1.000 \\
1 & DR & -0.0016 & 0.03939 & 0.995 \\
1 & LS & -0.0015 & 0.03964 & 1.002 \\
2 & RE & 0.0005 & 0.03365 & 1.000 \\
2 & DR & 0.0005 & 0.03257 & 0.968 \\
2 & LS & 0.0005 & 0.03262 & 0.970 \\
4 & RE & -0.0015 & 0.02782 & 1.000 \\
4 & DR & -0.0013 & 0.02516 & 0.904 \\
4 & LS & -0.0013 & 0.02518 & 0.905 \\
\end{longtable}

\clearpage
\textbf{Panel B. Interval coverage}

\begin{longtable}[]{@{}
  >{\raggedright\arraybackslash}p{(\columnwidth - 6\tabcolsep) * \real{0.1500}}
  >{\raggedright\arraybackslash}p{(\columnwidth - 6\tabcolsep) * \real{0.1900}}
  >{\raggedright\arraybackslash}p{(\columnwidth - 6\tabcolsep) * \real{0.3300}}
  >{\raggedright\arraybackslash}p{(\columnwidth - 6\tabcolsep) * \real{0.3300}}@{}}
\toprule\noalign{}
\begin{minipage}[b]{\linewidth}\raggedright
\textbf{γ}
\end{minipage} & \begin{minipage}[b]{\linewidth}\raggedright
\textbf{Fit}
\end{minipage} & \begin{minipage}[b]{\linewidth}\raggedright
\textbf{Plug-in coverage}
\end{minipage} & \begin{minipage}[b]{\linewidth}\raggedright
\textbf{mKH coverage}
\end{minipage} \\
\midrule\noalign{}
\endhead
\bottomrule\noalign{}
\endlastfoot
\multicolumn{4}{@{}>{\raggedright\arraybackslash}p{(\columnwidth - 6\tabcolsep) * \real{1.0000} + 6\tabcolsep}@{}}{%
\textbf{k = 20}} \\
0 & RE & 0.954 & 0.954 \\
0 & DR & 0.939 & 0.940 \\
0 & LS & 0.928 & 0.929 \\
1 & RE & 0.966 & 0.966 \\
1 & DR & 0.947 & 0.947 \\
1 & LS & 0.938 & 0.938 \\
2 & RE & 0.935 & 0.942 \\
2 & DR & 0.920 & 0.926 \\
2 & LS & 0.916 & 0.921 \\
4 & RE & 0.950 & 0.955 \\
4 & DR & 0.945 & 0.947 \\
4 & LS & 0.938 & 0.939 \\
\multicolumn{4}{@{}>{\raggedright\arraybackslash}p{(\columnwidth - 6\tabcolsep) * \real{1.0000} + 6\tabcolsep}@{}}{%
\textbf{k = 50}} \\
0 & RE & 0.943 & 0.944 \\
0 & DR & 0.935 & 0.936 \\
0 & LS & 0.934 & 0.934 \\
1 & RE & 0.961 & 0.961 \\
1 & DR & 0.955 & 0.956 \\
1 & LS & 0.952 & 0.953 \\
2 & RE & 0.950 & 0.951 \\
2 & DR & 0.946 & 0.946 \\
2 & LS & 0.945 & 0.945 \\
4 & RE & 0.948 & 0.949 \\
4 & DR & 0.940 & 0.944 \\
4 & LS & 0.940 & 0.944 \\
\end{longtable}

Figure 2 displays the corresponding RMSE ratios and makes the same point
graphically: performance is close to random effects around γ = 1, while
the clearest efficiency gain appears only under the strongest gradient.

Figure 2. RMSE ratio for constrained DR-Meta relative to conventional
random effects under τ₀² = 0.10. The empirically anchored γ = 1
condition lies near the no-gain boundary; the largest gain occurs only
at γ = 4.

\begin{center}
\includegraphics[width=0.96\linewidth,keepaspectratio]{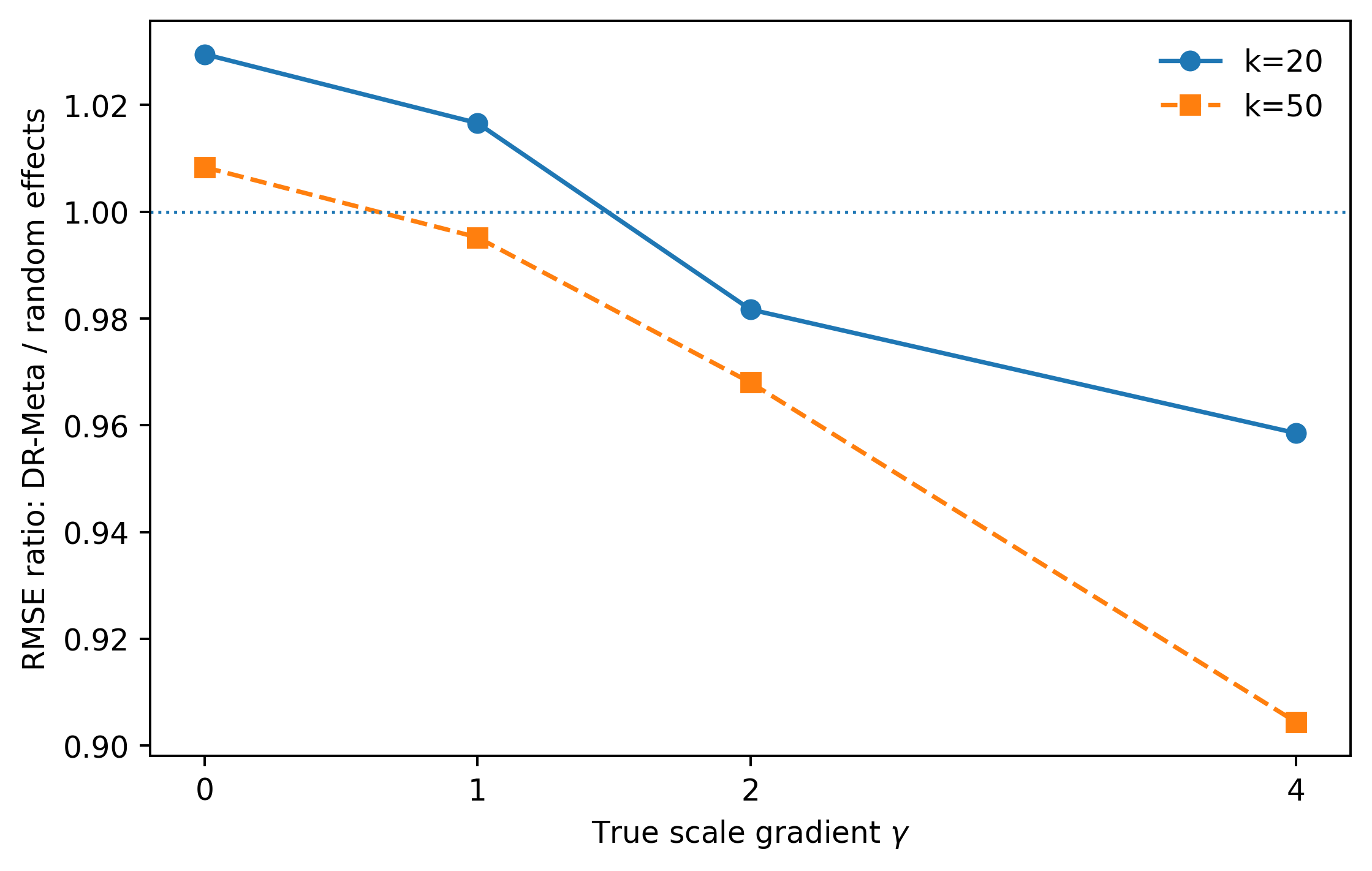}
\end{center}

\subsection{\texorpdfstring{\textbf{7.2 Coverage cost and the limited
role of
mKH}}{7.2 Coverage cost and the limited role of mKH}}\label{coverage-cost-and-the-limited-role-of-mkh}

Coverage is not merely noisy around the RE benchmark. Across many
conditions, constrained DR-Meta plug-in coverage is modestly lower than
RE coverage. This is a real inferential cost of estimating the scale
relation. The mKH adjustment usually adds only a few tenths to several
tenths of a percentage point and does not eliminate undercoverage in the
most difficult cells. For example, at k = 20, τ₀² = 0.10, γ = 2, DR-Meta
coverage rises from 0.920 to 0.926, while the corresponding RE coverage
is higher. At k = 50, τ₀² = 0.10, γ = 4, coverage rises from 0.940 to
0.944.

The appropriate conclusion is therefore not that mKH solves
scale-parameter uncertainty. It modestly improves a pragmatic interval
while profile or bootstrap procedures remain preferable when accurate
finite-sample coverage is central.

Figure 3 compares plug-in and modified Knapp-Hartung coverage for
constrained DR-Meta across the high-heterogeneity grid; the adjustment
is directionally helpful but does not remove the finite-sample deficit.

Figure 3. Empirical 95\% coverage for constrained DR-Meta under τ₀² =
0.10. Modified Knapp-Hartung (mKH) generally improves coverage slightly
but does not fully remove the finite-sample deficit.

\begin{center}
\includegraphics[width=0.96\linewidth,keepaspectratio]{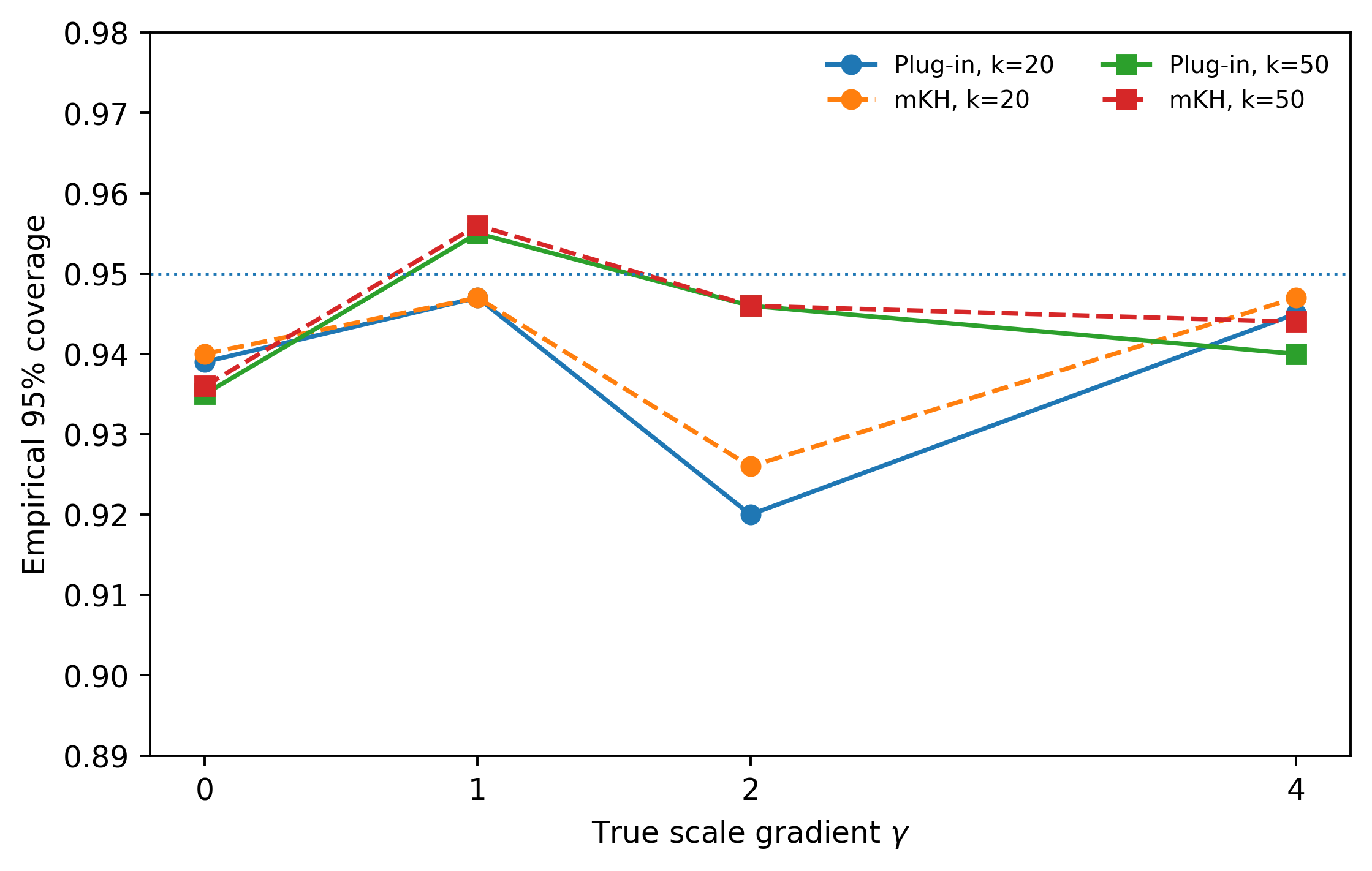}
\end{center}

\subsection{\texorpdfstring{\textbf{7.3 Boundary behavior under γ =
0}}{7.3 Boundary behavior under γ = 0}}\label{boundary-behavior-under-ux3b3-0}

Table 2 reports scale-gradient recovery together with lower- and
upper-bound behavior. The table should be read as an identification
diagnostic rather than as evidence that γ is accurately estimated in
every condition.

\clearpage
Table 2. Scale-gradient recovery and boundary behavior for constrained
DR-Meta

\begin{longtable}[]{@{}
  >{\raggedright\arraybackslash}p{(\columnwidth - 10\tabcolsep) * \real{0.1531}}
  >{\raggedright\arraybackslash}p{(\columnwidth - 10\tabcolsep) * \real{0.1106}}
  >{\raggedright\arraybackslash}p{(\columnwidth - 10\tabcolsep) * \real{0.1915}}
  >{\raggedright\arraybackslash}p{(\columnwidth - 10\tabcolsep) * \real{0.1915}}
  >{\raggedright\arraybackslash}p{(\columnwidth - 10\tabcolsep) * \real{0.1873}}
  >{\raggedright\arraybackslash}p{(\columnwidth - 10\tabcolsep) * \real{0.1660}}@{}}
\toprule\noalign{}
\begin{minipage}[b]{\linewidth}\raggedright
\textbf{τ₀²}
\end{minipage} & \begin{minipage}[b]{\linewidth}\raggedright
\textbf{γ}
\end{minipage} & \begin{minipage}[b]{\linewidth}\raggedright
\textbf{Mean \ensuremath{\hat{\gamma}}}
\end{minipage} & \begin{minipage}[b]{\linewidth}\raggedright
\textbf{Median \ensuremath{\hat{\gamma}}}
\end{minipage} & \begin{minipage}[b]{\linewidth}\raggedright
\textbf{Pr(\textless.001)}
\end{minipage} & \begin{minipage}[b]{\linewidth}\raggedright
\textbf{Pr(=8)}
\end{minipage} \\
\midrule\noalign{}
\endhead
\bottomrule\noalign{}
\endlastfoot
\multicolumn{6}{@{}>{\raggedright\arraybackslash}p{(\columnwidth - 10\tabcolsep) * \real{1.0000} + 10\tabcolsep}@{}}{%
\textbf{k = 20}} \\
0.02 & 0 & 1.417 & 0.197 & 0.476 & 0.039 \\
0.02 & 1 & 2.136 & 1.000† & 0.358 & 0.095 \\
0.02 & 2 & 2.432 & 1.000† & 0.333 & 0.125 \\
0.02 & 4 & 2.810 & 1.000† & 0.226 & 0.175 \\
0.10 & 0 & 0.854 & 0.065 & 0.488 & 0.003 \\
0.10 & 1 & 1.557 & 1.043 & 0.348 & 0.011 \\
0.10 & 2 & 2.573 & 2.166 & 0.189 & 0.052 \\
0.10 & 4 & 4.061 & 4.073 & 0.117 & 0.190 \\
\multicolumn{6}{@{}>{\raggedright\arraybackslash}p{(\columnwidth - 10\tabcolsep) * \real{1.0000} + 10\tabcolsep}@{}}{%
\textbf{k = 50}} \\
0.02 & 0 & 0.734 & 0.000 & 0.507 & 0.005 \\
0.02 & 1 & 1.642 & 1.075 & 0.314 & 0.034 \\
0.02 & 2 & 2.470 & 1.843 & 0.243 & 0.067 \\
0.02 & 4 & 3.366 & 2.384 & 0.162 & 0.189 \\
0.10 & 0 & 0.456 & 0.016 & 0.494 & 0.000 \\
0.10 & 1 & 1.139 & 0.969 & 0.209 & 0.000 \\
0.10 & 2 & 2.144 & 2.011 & 0.059 & 0.003 \\
0.10 & 4 & 4.244 & 4.017 & 0.020 & 0.094 \\
\end{longtable}

Recovery of γ depends strongly on the amount of scale information. The
three identical 1.000 medians in the k = 20, τ₀² = 0.02
positive-gradient rows are computational rather than substantive.
Inspection of the raw replication-level fits shows exact nonmovement at
the interior start γ = 1 in 5.8\%, 10.9\%, and 25.3\% of replications
when the true γ is 1, 2, and 4, respectively; the corresponding rate is
1.2\% under γ = 0. Thus the weakest-information cell contains a visibly
flat likelihood region in which the stopping rule can accept the initial
value. The same diagnostic is much smaller when heterogeneity is larger,
and pooled-location performance remains comparatively stable even when γ
is poorly identified. At k = 50 and τ₀² = 0.10, recovery is much
stronger: for true γ = 4, mean \ensuremath{\hat{\gamma}} = 4.244, median \ensuremath{\hat{\gamma}} = 4.017, only 2.0\%
of estimates are near zero, and 9.4\% reach the bound of 8. Mean
estimates should therefore be read jointly with medians, start-value
nonmovement, and both boundary frequencies.

\emph{† Median equals the optimizer\textquotesingle s interior starting
value γ = 1; the corresponding exact nonmovement rates are reported in
the text above. This mass is a computational diagnostic of weak
identification, not a recovery target.}

Under the boundary null γ = 0, the near-zero masses are 0.476, 0.488,
0.507, and 0.494 across the four k × τ₀² conditions. Their proximity to
one-half is qualitatively consistent with the familiar
single-boundary-parameter behavior described by Self and Liang11 and,
more directly, verifies that the direct-γ optimizer can actually attain
the null boundary. This is an implementation check, not a calibration
study of a universal mixture reference distribution.

Figure 4 plots mean and median \ensuremath{\hat{\gamma}} against the generating value. The
least-information panel visibly reflects the same optimizer-start
pile-up documented in Table 2; it should not be read as a recovery
plateau at γ = 1.

Figure 4. Recovery of the variance-gradient parameter

Note. The dashed 45-degree line denotes perfect recovery. Boundary
concentration and right skew are strongest when k or residual
heterogeneity is small. In the k = 20, τ₀² = 0.02 panel, exact
nonmovement at the interior start γ = 1 contributes to the median
pile-up documented in Table 2; this is a weak-identification diagnostic
rather than a substantive recovery pattern.

\begin{center}
\includegraphics[width=0.96\linewidth,keepaspectratio]{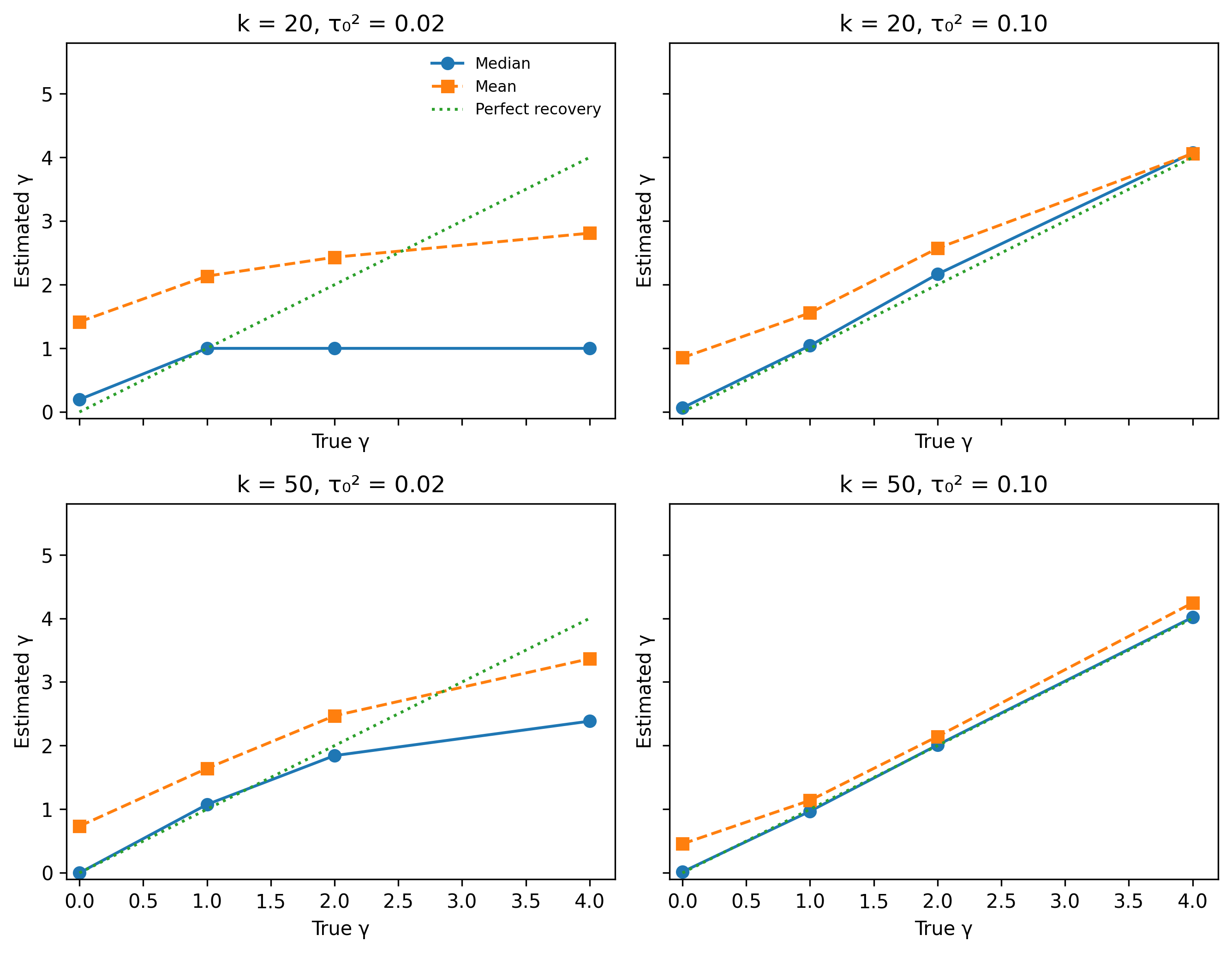}
\end{center}

\subsection{\texorpdfstring{\textbf{7.4 Misspecified scale
functions}}{7.4 Misspecified scale functions}}\label{misspecified-scale-functions}

Table 3 reports the three scale-misspecification mechanisms and shows
that the directional constraint either falls back toward random effects
or retains a gain only when the direction remains substantively
compatible.

\newpage

Table 3. Scale-model misspecification experiment (1,000 replications per
data-generating mechanism)

\textbf{Panel A. Point-estimation performance}

\begin{longtable}[]{@{}
  >{\raggedright\arraybackslash}p{(\columnwidth - 8\tabcolsep) * \real{0.3214}}
  >{\raggedright\arraybackslash}p{(\columnwidth - 8\tabcolsep) * \real{0.1339}}
  >{\raggedright\arraybackslash}p{(\columnwidth - 8\tabcolsep) * \real{0.1786}}
  >{\raggedright\arraybackslash}p{(\columnwidth - 8\tabcolsep) * \real{0.1786}}
  >{\raggedright\arraybackslash}p{(\columnwidth - 8\tabcolsep) * \real{0.1875}}@{}}
\toprule\noalign{}
\begin{minipage}[b]{\linewidth}\raggedright
\textbf{Pattern}
\end{minipage} & \begin{minipage}[b]{\linewidth}\raggedright
\textbf{Fit}
\end{minipage} & \begin{minipage}[b]{\linewidth}\raggedright
\textbf{Bias}
\end{minipage} & \begin{minipage}[b]{\linewidth}\raggedright
\textbf{RMSE}
\end{minipage} & \begin{minipage}[b]{\linewidth}\raggedright
\textbf{RMSE/RE}
\end{minipage} \\
\midrule\noalign{}
\endhead
\bottomrule\noalign{}
\endlastfoot
Increasing & RE & 0.0012 & 0.06343 & 1.000 \\
Increasing & DR & 0.0013 & 0.06371 & 1.004 \\
Increasing & LS & 0.0012 & 0.06299 & 0.993 \\
U-shaped & RE & -0.0017 & 0.03994 & 1.000 \\
U-shaped & DR & -0.0017 & 0.04031 & 1.009 \\
U-shaped & LS & -0.0018 & 0.04056 & 1.016 \\
Step decrease & RE & 0.0002 & 0.04833 & 1.000 \\
Step decrease & DR & -0.0001 & 0.04403 & 0.911 \\
Step decrease & LS & -0.0001 & 0.04408 & 0.912 \\
\end{longtable}

\textbf{Panel B. Coverage and scale-gradient behavior}

\begin{longtable}[]{@{}
  >{\raggedright\arraybackslash}p{(\columnwidth - 6\tabcolsep) * \real{0.3637}}
  >{\raggedright\arraybackslash}p{(\columnwidth - 6\tabcolsep) * \real{0.1515}}
  >{\raggedright\arraybackslash}p{(\columnwidth - 6\tabcolsep) * \real{0.2322}}
  >{\raggedright\arraybackslash}p{(\columnwidth - 6\tabcolsep) * \real{0.2526}}@{}}
\toprule\noalign{}
\begin{minipage}[b]{\linewidth}\raggedright
\textbf{Pattern}
\end{minipage} & \begin{minipage}[b]{\linewidth}\raggedright
\textbf{Fit}
\end{minipage} & \begin{minipage}[b]{\linewidth}\raggedright
\textbf{Coverage}
\end{minipage} & \begin{minipage}[b]{\linewidth}\raggedright
\textbf{Mean \ensuremath{\hat{\gamma}}}
\end{minipage} \\
\midrule\noalign{}
\endhead
\bottomrule\noalign{}
\endlastfoot
Increasing & RE & 0.941 & 0.000 \\
Increasing & DR & 0.940 & 0.104 \\
Increasing & LS & 0.935 & -1.034 \\
U-shaped & RE & 0.935 & 0.000 \\
U-shaped & DR & 0.927 & 0.566 \\
U-shaped & LS & 0.922 & -0.052 \\
Step decrease & RE & 0.941 & 0.000 \\
Step decrease & DR & 0.934 & 2.449 \\
Step decrease & LS & 0.934 & 2.439 \\
\end{longtable}

When heterogeneity increases with DR, constrained DR-Meta places γ at or
near zero in 82.7\% of replications and has essentially the same RMSE as
RE (ratio 1.004). The unrestricted model estimates a negative gradient
on average (mean \ensuremath{\hat{\gamma}} = -1.034) and is slightly more efficient than RE
(ratio 0.993). Under the U-shaped mechanism, constrained DR-Meta again
stays close to RE (RMSE ratio 1.009), with 52.2\% of γ estimates near
the lower boundary. Under the step-decrease mechanism, the direction is
correct even though the exponential shape is wrong; DR-Meta retains an
RMSE reduction of about 8.9\% (ratio 0.911), nearly identical to the
unrestricted LS fit.

These stress conditions directly evaluate the restriction rather than
merely acknowledging misspecification as a limitation. In the two
mechanisms that do not support a monotone decrease, the constrained
model degrades gracefully toward conventional random effects. When the
direction is correct but the shape is wrong, the exponential model can
still operate as a useful directional approximation. Coverage remains a
separate concern: for the step-decrease mechanism, plug-in coverage is
0.934 for DR-Meta versus 0.941 for RE.

Figure 5 plots the exact RMSE ratios reported in Table 3: 1.004 and
0.993 for the increasing mechanism, 1.009 and 1.016 for the U-shaped
mechanism, and 0.911 and 0.912 for the step-decrease mechanism for
constrained DR-Meta and unrestricted LS, respectively.

Figure 5. RMSE under scale-model misspecification, expressed relative to
conventional random effects. Ratios near 1 indicate graceful fallback;
the step-decrease mechanism retains a directional efficiency benefit
despite functional-form misspecification.

\begin{center}
\includegraphics[width=0.96\linewidth,keepaspectratio]{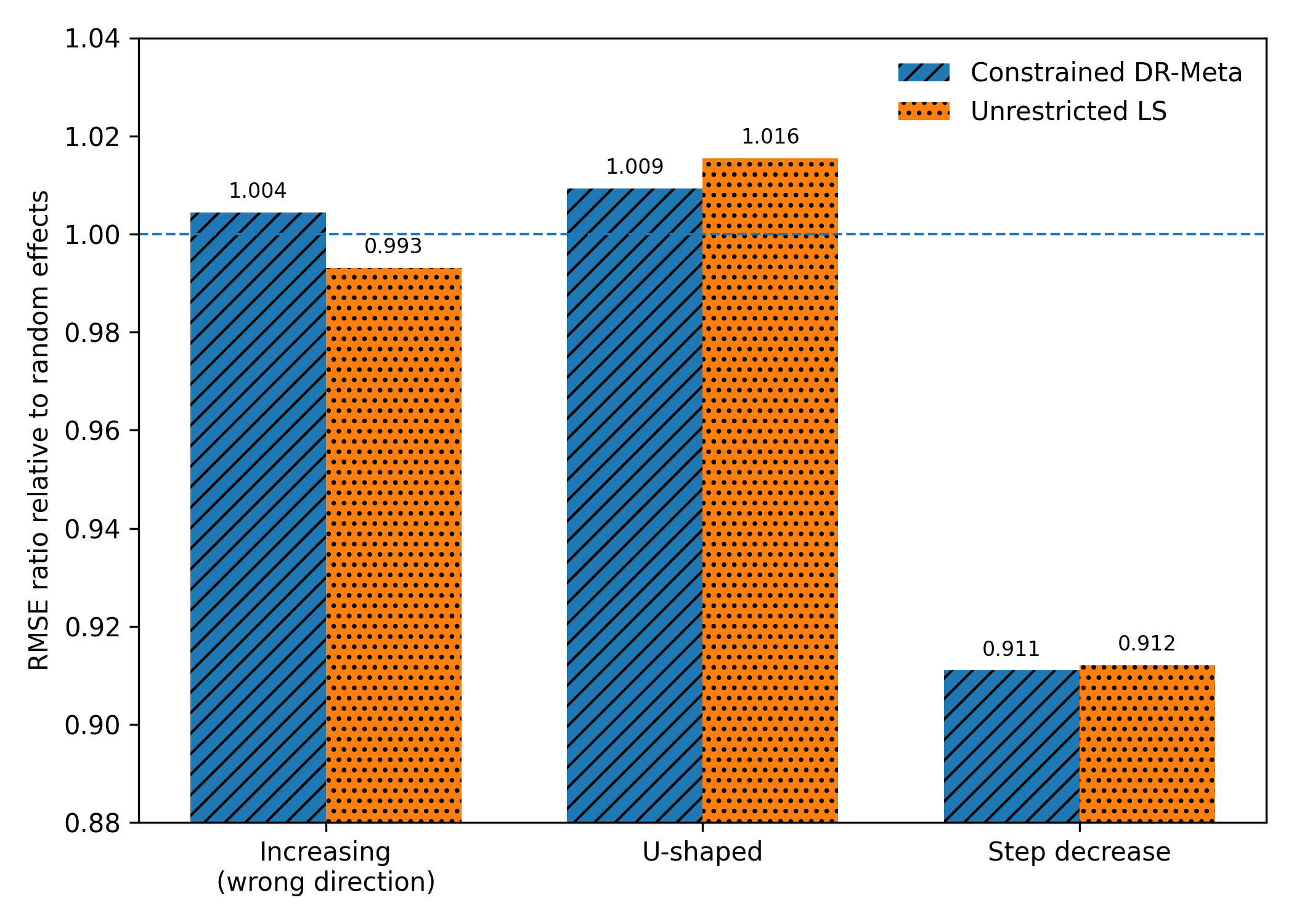}
\end{center}

\subsection{\texorpdfstring{\textbf{7.5 Sensitivity to the γ
optimization
bound}}{7.5 Sensitivity to the γ optimization bound}}\label{sensitivity-to-the-ux3b3-optimization-bound}

Table 4 reports the independent paired upper-bound sensitivity
experiment. Its γmax = 8 column is a comparator within that experiment,
not a duplicate of the main-run summaries in Tables 1 and 2. The
cross-run RMSE differences are smaller than approximately one combined
Monte Carlo standard error in all four γ = 4 conditions, so the modest
numerical discrepancies are consistent with Monte Carlo variation.
Widening the bound changes the scale-gradient estimate much more than
pooled-location RMSE.

Table 4. Sensitivity of γ recovery and pooled-estimate RMSE to the
optimization upper bound when true γ = 4

\begin{longtable}[]{@{}
  >{\raggedright\arraybackslash}p{(\columnwidth - 8\tabcolsep) * \real{0.1361}}
  >{\raggedright\arraybackslash}p{(\columnwidth - 8\tabcolsep) * \real{0.2179}}
  >{\raggedright\arraybackslash}p{(\columnwidth - 8\tabcolsep) * \real{0.2179}}
  >{\raggedright\arraybackslash}p{(\columnwidth - 8\tabcolsep) * \real{0.2141}}
  >{\raggedright\arraybackslash}p{(\columnwidth - 8\tabcolsep) * \real{0.2141}}@{}}
\toprule\noalign{}
\begin{minipage}[b]{\linewidth}\raggedright
\textbf{τ₀²}
\end{minipage} & \begin{minipage}[b]{\linewidth}\raggedright
\textbf{Mean \ensuremath{\hat{\gamma}}: 8}
\end{minipage} & \begin{minipage}[b]{\linewidth}\raggedright
\textbf{Mean \ensuremath{\hat{\gamma}}: 20}
\end{minipage} & \begin{minipage}[b]{\linewidth}\raggedright
\textbf{RMSE: 8}
\end{minipage} & \begin{minipage}[b]{\linewidth}\raggedright
\textbf{RMSE: 20}
\end{minipage} \\
\midrule\noalign{}
\endhead
\bottomrule\noalign{}
\endlastfoot
\multicolumn{5}{@{}>{\raggedright\arraybackslash}p{(\columnwidth - 8\tabcolsep) * \real{1.0000} + 8\tabcolsep}@{}}{%
\textbf{k = 20}} \\
0.02 & 2.643 & 3.879 & 0.03106 & 0.03111 \\
0.10 & 4.171 & 5.288 & 0.03962 & 0.03967 \\
\multicolumn{5}{@{}>{\raggedright\arraybackslash}p{(\columnwidth - 8\tabcolsep) * \real{1.0000} + 8\tabcolsep}@{}}{%
\textbf{k = 50}} \\
0.02 & 3.353 & 4.480 & 0.01901 & 0.01901 \\
0.10 & 4.165 & 4.494 & 0.02453 & 0.02456 \\
\end{longtable}

\emph{Note. Table 4 is a separate paired Monte Carlo sensitivity
experiment. Within each replicate, the same generated dataset is fit
under γmax = 8 and γmax = 20, but these datasets are independent of the
main factorial run used for Tables 1 and 2. Accordingly, the γmax = 8
marginal summaries are not expected to reproduce Tables 1 and 2 exactly;
the small cross-table differences are Monte Carlo variation, while the
paired within-table comparison isolates the effect of widening the
optimization bound.}

Widening the upper bound substantially changes mean \ensuremath{\hat{\gamma}} and moves many
estimates away from the artificial value 8. Under the {[}0,8{]} fit, the
proportions at the upper bound are 0.163, 0.211, 0.174, and 0.083 for
(k, τ₀²) = (20,.02), (20,.10), (50,.02), and (50,.10); under {[}0,20{]},
the corresponding proportions at 20 fall to 0.072, 0.048, 0.055, and
0.006. Yet pooled-estimate RMSE changes by less than about two-tenths of
one percent in every condition. The practical implication is asymmetric:
the fitted location can be stable even when γ itself is poorly
identified. A reported γ near an optimization limit should therefore
trigger a wider-bound sensitivity analysis, but it does not
automatically invalidate the pooled location.

\subsection{\texorpdfstring{\textbf{7.6 Score quality and mean
misspecification}}{7.6 Score quality and mean misspecification}}\label{score-quality-and-mean-misspecification}

The score-quality experiment separates degradation of the design
moderator from misspecification of the statistical model. With the exact
score, DR-Meta has RMSE = 0.02491 compared with 0.02700 for RE, a 7.8\%
reduction, while coverage is 0.948. Adding independent score error
increases RMSE to 0.02525 but retains a 6.5\% reduction relative to RE.
Coarsening DR to three levels gives RMSE = 0.02534, a 6.2\% reduction.
The clearer consequence of score degradation is attenuation of the scale
gradient: mean \ensuremath{\hat{\gamma}} falls from 4.198 with the exact score to 3.058 with
noise and 2.627 after coarsening.

\subsubsection{\texorpdfstring{\textbf{7.6.1 DR-score
quality}}{7.6.1 DR-score quality}}\label{dr-score-quality}

Table 5 reports the paired score-quality experiment, separating
attenuation of γ from the more modest change in pooled-effect RMSE.

\newpage

Table 5. Paired design-robustness score-quality experiment

\begin{samepage}
\textbf{Panel A. Pooled-effect performance}

\begin{center}
{\small
\begin{tabular}{lllll}
\toprule
\textbf{Score} & \textbf{Fit} & \textbf{Bias} & \textbf{RMSE} & \textbf{Coverage} \\
\midrule
Exact & DR & -.0010 & .02491 & .948 \\
Exact & RE & -.0007 & .02700 & .943 \\
Noisy & DR & -.0009 & .02525 & .944 \\
Noisy & RE & -.0007 & .02700 & .943 \\
3 levels & DR & -.0008 & .02534 & .941 \\
3 levels & RE & -.0007 & .02700 & .943 \\
\bottomrule
\end{tabular}
}
\end{center}

\textbf{Panel B. Scale-gradient summaries}

\begin{center}
{\small
\begin{tabular}{llll}
\toprule
\textbf{Score} & \textbf{Fit} & \textbf{Mean \ensuremath{\hat{\gamma}}} & \textbf{Median \ensuremath{\hat{\gamma}}} \\
\midrule
Exact & DR & 4.198 & 4.042 \\
Exact & RE & --- & --- \\
Noisy & DR & 3.058 & 2.842 \\
Noisy & RE & --- & --- \\
3 levels & DR & 2.627 & 2.321 \\
3 levels & RE & --- & --- \\
\bottomrule
\end{tabular}
}
\end{center}
\end{samepage}

Figure 6 visualizes the score-quality result: noisy and coarsened scores
weaken the estimated gradient but retain most of the point-estimation
advantage in this deliberately strong-gradient condition.

Figure 6. RMSE under exact, noisy, and coarsened design-robustness
scores

\begin{center}
\includegraphics[width=0.74\linewidth,keepaspectratio]{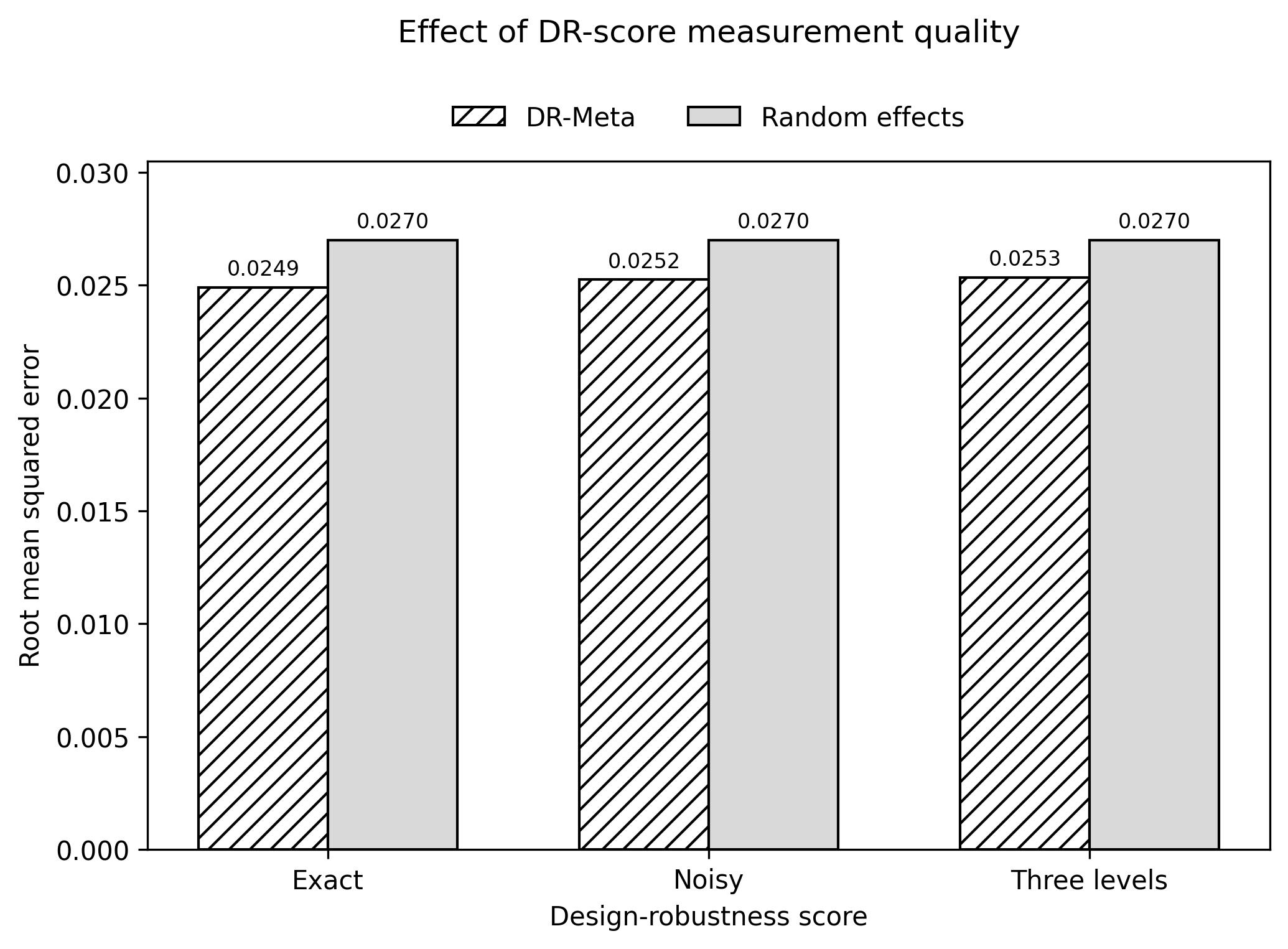}
\end{center}

Note. The same 1,000 generated datasets are used across score
conditions. RE is identical across conditions by construction; γ
summaries apply only to DR-Meta.

\subsubsection{\texorpdfstring{\textbf{7.6.2 Design-linked mean
misspecification}}{7.6.2 Design-linked mean misspecification}}\label{design-linked-mean-misspecification}

Scale reweighting cannot substitute for a location model. When δ = 0, RE
and scale-only DR-Meta are essentially unbiased, and DR-Meta has
slightly lower RMSE. With δ = 0.10, RE bias is 0.0502 and DR-Meta bias
is 0.0430, whereas MR and JLS biases are 0.0052 and 0.0040. Coverage is
0.620 for RE, 0.651 for DR-Meta, 0.977 for MR, and 0.952 for JLS. At δ =
0.20, bias increases to 0.1001 for RE and 0.0840 for DR-Meta, while MR
and JLS remain close to zero (-0.0024 and -0.0019); their coverages are
0.965 and 0.948 compared with 0.094 and 0.169 for RE and DR-Meta.

Thus, scale reweighting moves the pooled estimate toward the high-DR
target and attenuates the imposed mean distortion by roughly 14\%-16\%,
but it does not repair the misspecified mean model. At δ = 0.20, JLS
also has lower RMSE than MR (0.06371 versus 0.06737), showing that scale
modeling can add efficiency after the relevant location gradient has
been represented. The mean \ensuremath{\hat{\gamma}} sequence for scale-only DR-Meta (3.126,
3.239, 3.106) is nonmonotone; it is therefore reported descriptively
rather than interpreted as systematic leakage from location to scale.

Table 6 reports the mean-misspecification experiment and makes the
location-scale division of labor explicit: scale reweighting attenuates
but does not remove a systematic design-linked mean shift.

Table 6. Estimation of the high-DR target under design-linked mean
shifts

\begin{longtable}[]{@{}
  >{\raggedright\arraybackslash}p{(\columnwidth - 10\tabcolsep) * \real{0.1247}}
  >{\raggedright\arraybackslash}p{(\columnwidth - 10\tabcolsep) * \real{0.1291}}
  >{\raggedright\arraybackslash}p{(\columnwidth - 10\tabcolsep) * \real{0.1849}}
  >{\raggedright\arraybackslash}p{(\columnwidth - 10\tabcolsep) * \real{0.1849}}
  >{\raggedright\arraybackslash}p{(\columnwidth - 10\tabcolsep) * \real{0.1613}}
  >{\raggedright\arraybackslash}p{(\columnwidth - 10\tabcolsep) * \real{0.2151}}@{}}
\toprule\noalign{}
\begin{minipage}[b]{\linewidth}\raggedright
\textbf{δ}
\end{minipage} & \begin{minipage}[b]{\linewidth}\raggedright
\textbf{Fit}
\end{minipage} & \begin{minipage}[b]{\linewidth}\raggedright
\textbf{Bias}
\end{minipage} & \begin{minipage}[b]{\linewidth}\raggedright
\textbf{RMSE}
\end{minipage} & \begin{minipage}[b]{\linewidth}\raggedright
\textbf{Cov}
\end{minipage} & \begin{minipage}[b]{\linewidth}\raggedright
\textbf{Mean \ensuremath{\hat{\gamma}}}
\end{minipage} \\
\midrule\noalign{}
\endhead
\bottomrule\noalign{}
\endlastfoot
0.0 & RE & -.0001 & .02881 & .955 & --- \\
0.0 & DR & -.0001 & .02698 & .948 & 3.126 \\
0.0 & MR & -.0024 & .06638 & .976 & --- \\
0.0 & JLS & -.0038 & .06174 & .954 & 3.133 \\
0.1 & RE & .0502 & .05789 & .620 & --- \\
0.1 & DR & .0430 & .05103 & .651 & 3.239 \\
0.1 & MR & .0052 & .06496 & .977 & --- \\
0.1 & JLS & .0040 & .06077 & .952 & 3.261 \\
0.2 & RE & .1001 & .10480 & .094 & --- \\
0.2 & DR & .0840 & .08919 & .169 & 3.106 \\
0.2 & MR & -.0024 & .06737 & .965 & --- \\
0.2 & JLS & -.0019 & .06371 & .948 & 3.120 \\
\end{longtable}

Figure 7 visualizes the same result, with the mean-only and joint
location models remaining near the high-DR target as the imposed mean
gradient increases.

Figure 7. Bias for the high-DR target under design-linked mean
misspecification

\begin{center}
\includegraphics[width=0.90\linewidth,keepaspectratio]{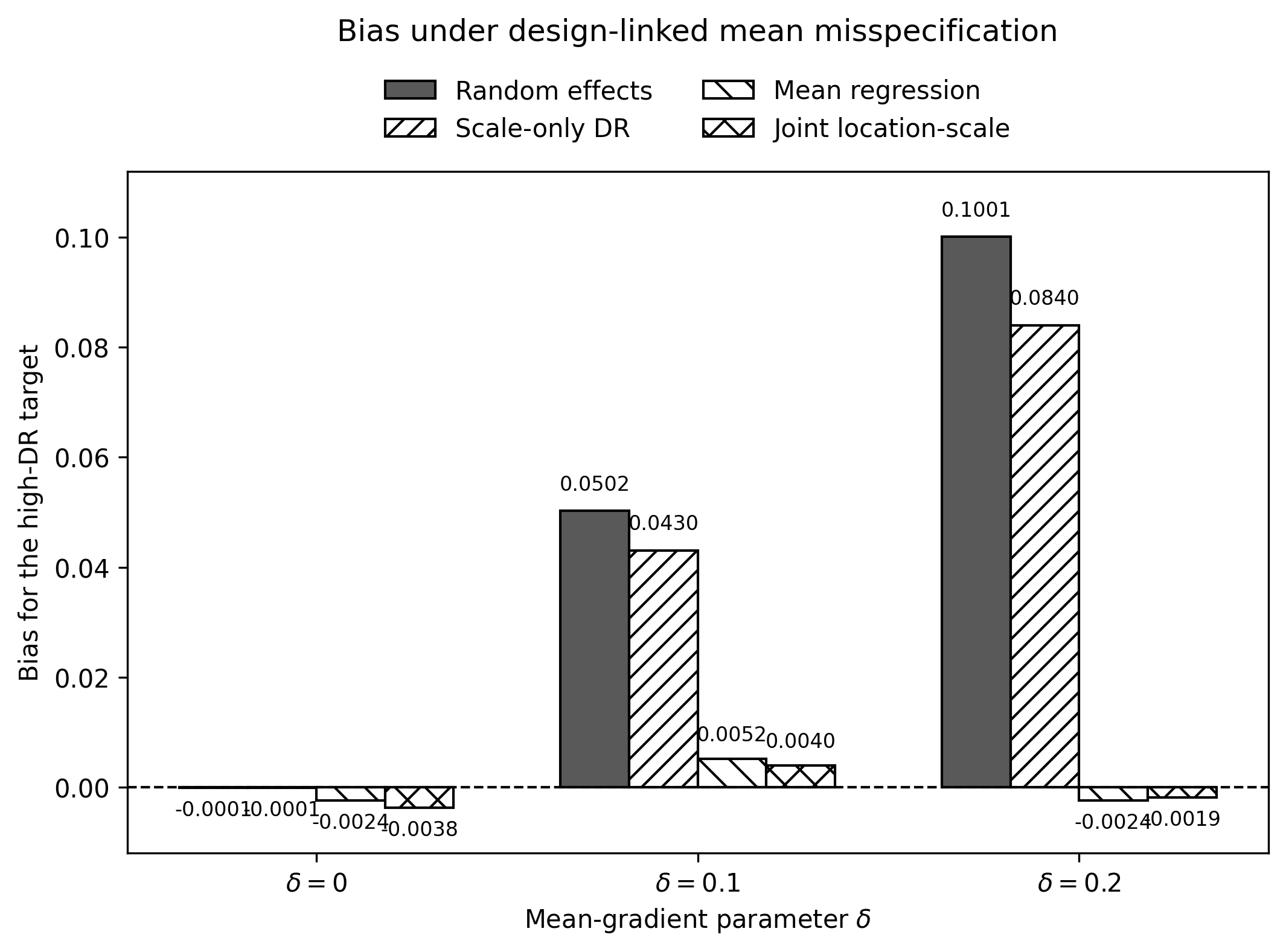}
\end{center}

Note. Scale-only DR-Meta can attenuate design-linked distortion through
reweighting, but an explicit location model is required to represent a
systematic design-linked shift in the mean.

\section{\texorpdfstring{\textbf{8. Empirical Illustration:
Outcome-Separated Design Quality in a PSM/PSW Evidence
Base}}{8. Empirical Illustration: Outcome-Separated Design Quality in a PSM/PSW Evidence Base}}\label{empirical-illustration-outcome-separated-design-quality-in-a-psmpsw-evidence-base}

\subsection{\texorpdfstring{\textbf{8.1 Data and DR
construction}}{8.1 Data and DR construction}}\label{data-and-dr-construction}

The empirical illustration uses an 18-study evidence base of
propensity-score matching or weighting comparisons of online/hybrid
versus face-to-face instruction. The purpose is methodological rather
than substantive: it asks whether an outcome-separated design score can
be constructed and whether the constrained scale relation is actually
supported in a real synthesis.

The design-quality score ranges from 0 to 7 and gives one point each for
theoretically grounded covariate selection, validated propensity-score
estimation, an appropriate matching or weighting algorithm, reported
standardized-mean-difference balance diagnostics, demonstrated overlap,
caliper or trimming use, and a sensitivity analysis. These criteria are
grounded in propensity-score design principles.\textsuperscript{16,17}

For DR-Meta the score is normalized as DR = quality/7, preserving the
prespecified theoretical scale rather than stretching the observed
sample to fill {[}0,1{]}. The 18 studies occupy observed support from
4/7 to 6/7. Effect sizes are Cohen's d and the reported study-level
standard errors are used to form sampling variances. Appendix C lists
the complete study-level input used here.

\subsection{\texorpdfstring{\textbf{8.2
Results}}{8.2 Results}}\label{results}

Table 7 reports the empirical illustration across conventional random
effects, constrained and unrestricted scale fits, and location models.

Table 7. Empirical illustration using DR = PSM/PSW design-quality score
divided by 7. For MR and joint location-scale models, the location
estimate is the fitted intercept at DR = 1.

\begin{longtable}[]{@{}
  >{\raggedright\arraybackslash}p{(\columnwidth - 8\tabcolsep) * \real{0.3360}}
  >{\raggedright\arraybackslash}p{(\columnwidth - 8\tabcolsep) * \real{0.1779}}
  >{\raggedright\arraybackslash}p{(\columnwidth - 8\tabcolsep) * \real{0.1581}}
  >{\raggedright\arraybackslash}p{(\columnwidth - 8\tabcolsep) * \real{0.1739}}
  >{\raggedright\arraybackslash}p{(\columnwidth - 8\tabcolsep) * \real{0.1541}}@{}}
\toprule\noalign{}
\begin{minipage}[b]{\linewidth}\raggedright
\textbf{Fit}
\end{minipage} & \begin{minipage}[b]{\linewidth}\raggedright
\textbf{Location}
\end{minipage} & \begin{minipage}[b]{\linewidth}\raggedright
\textbf{SE}
\end{minipage} & \begin{minipage}[b]{\linewidth}\raggedright
\textbf{τ₀²}
\end{minipage} & \begin{minipage}[b]{\linewidth}\raggedright
\textbf{\ensuremath{\hat{\gamma}}}
\end{minipage} \\
\midrule\noalign{}
\endhead
\bottomrule\noalign{}
\endlastfoot
RE & -0.107 & 0.101 & 0.175 & 0.000 \\
Constrained DR & -0.107 & 0.101 & 0.175 & 0.000 \\
Unrestricted LS & -0.114 & 0.100 & 0.071 & -1.249 \\
Mean-only MR & 0.059 & 0.299 & 0.182 & 0.000 \\
Joint location-scale & 0.155 & 0.277 & 1.055 & 2.443 \\
\end{longtable}

The key result is diagnostic. Constrained DR-Meta lands exactly at \ensuremath{\hat{\gamma}} = 0
and is numerically identical to conventional random effects (\ensuremath{\hat{\mu}} = -0.107,
SE = 0.101, \ensuremath{\hat{\tau}}² = 0.175). The unrestricted scale model estimates \ensuremath{\hat{\gamma}} =
-1.249, indicating that the best-fitting unconstrained variance function
increases rather than decreases over the observed DR range. Its fitted
residual-variance ratio from DR = 4/7 to 6/7 is approximately 1.43, the
opposite of the prespecified directional hypothesis.

Accordingly, the empirical application does not ``validate'' DR-Meta by
finding a favorable gradient. It demonstrates the intended guardrail:
when the design dimension does not support declining heterogeneity, the
constrained estimator falls back to conventional random effects while
the unrestricted fit signals directional disagreement.

A mean-only meta-regression gives a fitted DR = 1 intercept of 0.059 (SE
= 0.299), and the joint location-scale model gives 0.155 (SE = 0.277)
with \ensuremath{\hat{\gamma}} = 2.443. Both location intercepts are very imprecise because the
observed data occupy only 4/7 to 6/7 of the theoretical DR range. The
joint model's τ₀² = 1.055 is an even more direct extrapolation: τ₀² is
defined at DR = 0, which lies well outside observed support. Over the
observed range, that fitted scale corresponds instead to τ²(4/7) ≈ 0.261
and τ²(6/7) ≈ 0.130. Neither the DR = 1 location intercept nor the DR =
0 scale intercept is therefore used as a primary substantive quantity.

A complete leave-one-out analysis shows that the directional conclusion
is not driven by a single study. Across all 18 deletions, the
constrained estimate remains exactly \ensuremath{\hat{\gamma}} = 0 in 15 fits; it becomes only
0.458, 0.219, and 0.090 when Askew et al., Babcock and Georgiou, or
Hobbs is removed, respectively. In the same 15 deletions for which the
constrained fit remains at zero, the unrestricted scale estimate remains
negative. Conway et al. is influential for location and overall
heterogeneity: removing it changes the conventional RE estimate from
-0.107 (\ensuremath{\hat{\tau}}² = 0.175) to -0.038 (\ensuremath{\hat{\tau}}² = 0.105). Yet the constrained fit
still has \ensuremath{\hat{\gamma}} = 0 and the unrestricted fit remains negative (\ensuremath{\hat{\gamma}} ≈ -1.165).
Thus the extreme Conway estimate materially affects the pooled location
but does not create the conclusion that the prespecified
declining-heterogeneity direction is unsupported.

The empirical example should therefore be read as a falsification
opportunity rather than a showcase. The score is constructed from
design/process criteria without using the realized effect estimates,
satisfying the primary separation rule, yet the observed evidence does
not support the prespecified declining-heterogeneity direction. That is
precisely the behavior a constrained model should permit. Because the
illustration reuses an independently assembled evidence base, it is not
presented as a substantive review of online learning; its purpose is to
demonstrate score construction, directional diagnosis, and the
distinction between location and scale.

Figure 8 shows the empirical evidence directly: the constrained
scale-only fit coincides with random effects at γ = 0, while the
unrestricted diagnostic points toward increasing rather than decreasing
residual heterogeneity with DR.

Figure 8. Study effects in the 18-study PSM/PSW illustration by
normalized design robustness. Circle size reflects study precision. The
constrained scale-only DR-Meta fit coincides with conventional random
effects at γ = 0; the dotted line shows the highly uncertain mean-only
location trend.

\begin{center}
\includegraphics[width=0.96\linewidth,keepaspectratio]{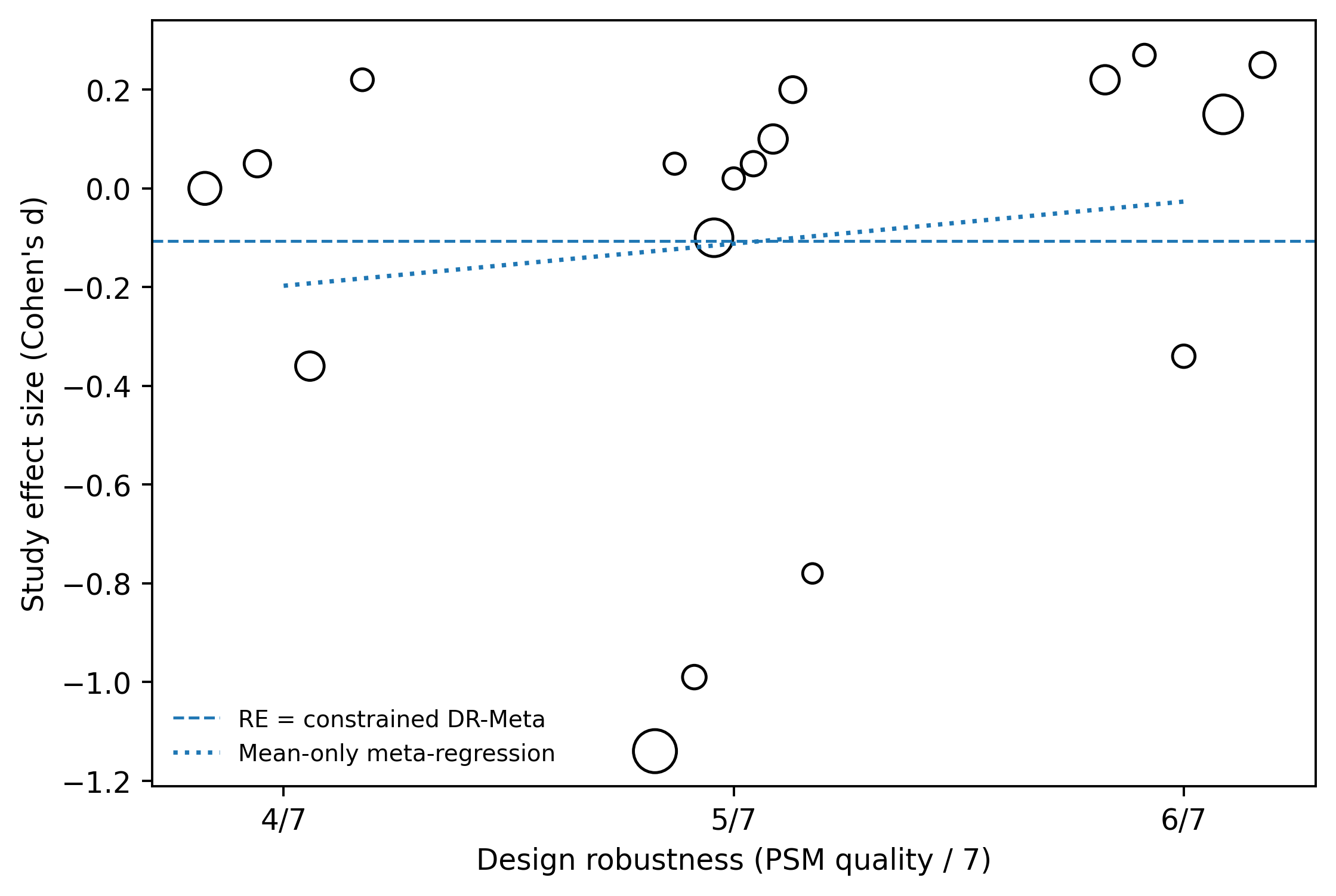}
\end{center}

\section{\texorpdfstring{\textbf{9. Computational Validation and
Reproducibility}}{9. Computational Validation and Reproducibility}}\label{computational-validation-and-reproducibility}

\subsection{\texorpdfstring{\textbf{9.1 Software
implementation}}{9.1 Software implementation}}\label{software-implementation}

The public R implementation is drmeta version 0.2.2, released on CRAN on
August 24, 2026 and preserved as the versioned GitHub release v0.2.2.
The package supports conventional random effects through γ fixed at 0,
constrained DR-Meta, a signed unrestricted scale diagnostic, location
moderators, and the joint location-scale specification; companion
functions provide scale prediction, observed-support variance contrasts,
heterogeneity diagnostics, leave-one-out analyses, and a
parametric-bootstrap utility. The R package is the computational source for Tables 5 and 6 and the
public implementation of the model. For Tables 1-4, B1, and the empirical
reference calculations, the manuscript retains the exact archived
fit-level outputs used for the reported numerical results; supplied R
scripts reproduce the summaries and figures from those archived
artifacts.\textsuperscript{15}

For readers already using metafor, the unrestricted signed scale model
can be fit directly with rma(..., scale = \textasciitilde{} DR) under
the default log link. The additional role of drmeta is the one-sided γ ≥
0 parameterization and the associated boundary, scale-contrast, and
diagnostic workflow; the general location-scale likelihood itself is not
package-specific.

This division of labor is intentional and is recorded in the bundle
provenance map. The R/drmeta 0.2.2 run uses master seed 20260806 for the
package-based score-quality and mean-misspecification experiments. The
archive supplies the exact fit-level numerical artifacts used for the
expanded main grid, mKH comparisons, deliberate scale-function
violations, wider-bound experiment, and optimizer diagnostics. Fresh R
reruns of the declared simulation designs are expected to agree in
operating characteristics rather than reproduce every archived last
digit when the original random stream is unavailable.

\subsection{\texorpdfstring{\textbf{9.2 Independent numerical
checks}}{9.2 Independent numerical checks}}\label{independent-numerical-checks}

The profiled REML objective was checked with the public R/drmeta
implementation and an independently coded reference implementation
during development. Structural checks confirm exact nesting at γ = 0, and the empirical Table 7 calculations were independently
re-derived from the model equations. In development comparisons on
shared datasets, conventional RE and mean-only meta-regression estimates
agreed to numerical tolerance and pooled-location differences between
scale implementations were negligible when the likelihood was well
identified. In small-k, low-heterogeneity conditions, scale parameters
can select near-equivalent optima; this is treated as weak
identification rather than hidden by averaging across engines.

Separate numerical checks also verify the sign of the weight derivative
in Proposition 2, the fixed-effect upper bound in Proposition 3, the
pseudo-true target in Appendix A4, and affine score-rescaling behavior
in Appendix A5. These checks supplement rather than replace the
algebraic proofs.

\subsection{\texorpdfstring{\textbf{9.3 Reproducibility
files}}{9.3 Reproducibility files}}\label{reproducibility-files}

The reproducibility bundle contains: (1) exact archived raw and summary
outputs for Tables 1-4 and B1; (2) R/drmeta 0.2.2 package scripts and
recovered package-run summary files for Tables 5 and 6; (3) the
full-precision 18-study empirical input, Table 7 results, and all 18
leave-one-out fits; (4) R scripts that recompute manuscript summaries
from the archived raw outputs and regenerate Figures 1-8; (5) corrected
signed-scale near-zero diagnostics for the unrestricted LS stress fits;
and (6) a run manifest, a detailed provenance map, and SHA-256 checksums.
The exact original random stream for every copied stress/bound artifact
was not preserved when those analyses were first expanded, so the
archived raw fit-level output is the exact numerical target; fresh
fixed-seed R reruns of the declared designs are expected to agree up to
Monte Carlo error rather than byte-for-byte. The manuscript-specific reproducibility bundle is included with this arXiv submission as ancillary material.

\section{\texorpdfstring{\textbf{10. Practical Workflow and Reporting
Standards}}{10. Practical Workflow and Reporting Standards}}\label{practical-workflow-and-reporting-standards}

\subsection{\texorpdfstring{\textbf{10.1 Before fitting the
model}}{10.1 Before fitting the model}}\label{before-fitting-the-model}

\begin{itemize}
\item
  Define one common effect-size metric, verify every transformation and
  sampling variance, state the synthesis estimand, and decide whether a
  constant conditional mean is scientifically plausible before
  constructing DR.
\item
  Specify and document each DR component, coding rule, and any component
  weight before looking at the meta-analytic effect estimates whenever
  the analysis is intended to be confirmatory.
\item
  Document the theoretical DR scale, empirical range, distribution, and
  number of distinct values. Do not estimate a scale gradient from
  nearly constant DR.
\item
  Decide in advance which observed-support contrast will be used for Rτ
  or Aτ, and avoid automatic extrapolation from DR = 0 to DR = 1 when
  those endpoints are unsupported.
\end{itemize}

For the primary analysis, do not use yᵢ, its standard error, p value,
confidence interval, or outcome-dependent robustness diagnostics in
constructing DR. If outcome-dependent information is scientifically
useful, report that score only as a clearly labeled exploratory
alternative.

Inspect whether DR is associated with sampling precision or other study
characteristics before interpreting the fitted scale relation. A design
score can be outcome-separated and still be confounded with study size,
setting, intervention class, or measurement quality.

\subsection{\texorpdfstring{\textbf{10.2 Minimum comparison
set}}{10.2 Minimum comparison set}}\label{minimum-comparison-set}

A primary DR-Meta analysis should not be reported alone. At minimum,
report conventional random effects, constrained DR-Meta, and an
unrestricted scale fit. If design-linked mean shifts are plausible, add
a location-only meta-regression and a joint location-scale model. This
comparison set separates three questions: whether the scale relation
exists, whether its direction agrees with the constraint, and whether
the mean itself changes with DR.

\subsection{\texorpdfstring{\textbf{10.3 Sensitivity
analyses}}{10.3 Sensitivity analyses}}\label{sensitivity-analyses}

\begin{itemize}
\item
  Widen γ bounds when estimates accumulate at the optimization limit.
\item
  Refit after influential-study deletion and under defensible
  alternative DR constructions.
\item
  Compare plug-in and small-sample or resampling intervals for location
  parameters.
\item
  If the unrestricted scale coefficient reverses direction, treat the
  constrained γ = 0 result as substantive evidence against the
  prespecified scale hypothesis rather than as a failed optimization.
\item
  Avoid interpreting DR-weight changes as removal of confounding or
  study bias.
\end{itemize}

Use leave-one-out and influence diagnostics for both location and scale
components. A study can be unremarkable for the pooled mean yet decisive
for the estimated variance gradient.

Report primary and exploratory score constructions separately; examine
component-wise scores, reasonable coarsenings, and uncertainty in DR
where feasible. If substantive conclusions depend on one arbitrary score
mapping, the evidence for a design-indexed scale process is weak.

\subsection{\texorpdfstring{\textbf{10.4 Suggested reporting
language}}{10.4 Suggested reporting language}}\label{suggested-reporting-language}

A concise report should state that DR-Meta evaluates a directional
residual-heterogeneity specification; give the observed DR support;
report \ensuremath{\hat{\gamma}}, boundary status, and an observed-support variance contrast;
compare RE, constrained, and unrestricted fits; and state explicitly
whether a location model was required. If \ensuremath{\hat{\gamma}} = 0, report that the
constrained model nested at random effects. If \ensuremath{\hat{\gamma}} is large but unstable
to the optimization bound, emphasize the stability or instability of the
pooled location separately from the scale parameter.

\section{\texorpdfstring{\textbf{11.
Discussion}}{11. Discussion}}\label{discussion}

\subsection{\texorpdfstring{\textbf{11.1 What the evidence
supports}}{11.1 What the evidence supports}}\label{what-the-evidence-supports}

The evidence supports a deliberately limited claim for DR-Meta. The
method is not a general efficiency improvement over random effects.
Around an empirically anchored gradient near γ = 1, gains are negligible
and can be slightly negative in smaller meta-analyses. Useful
point-estimation gains appear when the residual scale gradient is
stronger and residual heterogeneity is substantial. The largest
improvement belongs to an extreme γ = 4 cell and should be presented as
such.

The misspecification experiment is more informative than the maximum
gain. When the directional constraint is wrong, the constrained fit
usually collapses toward γ = 0 and behaves like random effects. Under a
nonmonotone U-shape it again stays near the conventional model. When the
direction is correct but the exponential shape is wrong, the model can
still extract an efficiency benefit. This pattern is exactly what a
constrained working model should do: use the directional structure when
the data support it and retreat when they do not.

\subsection{\texorpdfstring{\textbf{11.2 Efficiency is purchased with
inferential
uncertainty}}{11.2 Efficiency is purchased with inferential uncertainty}}\label{efficiency-is-purchased-with-inferential-uncertainty}

The coverage results also impose a real qualification. DR-Meta plug-in
intervals are often modestly less conservative than conventional RE
intervals because they condition on estimated scale parameters. The mKH
comparison is helpful but not decisive; it generally improves coverage
only slightly. Users should therefore separate point-estimation
efficiency from interval calibration and use bootstrap or profile
methods when finite-sample inference is central.

\subsection{\texorpdfstring{\textbf{11.3 The empirical illustration is a
falsification
opportunity}}{11.3 The empirical illustration is a falsification opportunity}}\label{the-empirical-illustration-is-a-falsification-opportunity}

The 18-study PSM/PSW example is intentionally useful because it does not
cooperate with the proposed direction. The constrained fit lands at γ =
0, while the unrestricted fit points toward increasing residual variance
with higher design scores. A method paper is stronger when its empirical
example can reject the favored structural assumption. Here the
constraint functions as a falsifiable statement rather than as a
mechanism for mechanically rewarding higher-scored studies.

\subsection{\texorpdfstring{\textbf{11.4 Scale identification can be
weak while location is
stable}}{11.4 Scale identification can be weak while location is stable}}\label{scale-identification-can-be-weak-while-location-is-stable}

The γ-bound experiment shows that scale-gradient estimation can be much
less stable than the pooled location. Extending the upper bound from 8
to 20 materially changes mean \ensuremath{\hat{\gamma}} under strong-gradient conditions, but
pooled RMSE barely changes. This supports a reporting hierarchy: first
report the location estimate and its uncertainty, then the scale
relation and its bound sensitivity, and only then interpret γ or
attenuation contrasts.

This separation should guide interpretation of raw γ. Because γ changes
under rescaling of DR, its numerical magnitude is not an invariant
measure of design sensitivity. Reporting the observed DR support
together with an observed-support variance ratio Rτ or attenuation Aτ
gives a more interpretable description of the fitted scale relation.
Appendix A5 shows why these fitted contrasts are invariant to a positive
affine re-expression of the score when the model is transformed
coherently.

\subsection{\texorpdfstring{\textbf{11.5 Relation to general
location-scale
models}}{11.5 Relation to general location-scale models}}\label{relation-to-general-location-scale-models}

The statistical parent model and much of its inferential machinery are
established; DR-Meta does not claim priority for heterogeneity
moderation or location-scale meta-analysis.\textsuperscript{7,8}

Its narrower contribution is to combine a prespecified outcome-separated
design index, a directional scale restriction, exact nesting at random
effects, formal weight properties, observed-support contrasts, and
explicit constrained-versus-unrestricted diagnostics. Kuper and
colleagues demonstrate why study characteristics and risk-of-bias
information can be empirically relevant for residual heterogeneity;
DR-Meta asks how a prespecified directional design hypothesis should be
encoded and challenged.\textsuperscript{13}

\subsection{\texorpdfstring{\textbf{11.6
Limitations}}{11.6 Limitations}}\label{limitations}

Several limitations remain. First, DR is an analyst-constructed
study-level moderator and can contain judgment or measurement error even
when outcome separation is respected. The separation condition prevents
direct outcome leakage but does not establish causal exogeneity. Second,
study-level relations between design features and effect dispersion are
ecological across studies and should not be interpreted as within-study
causal effects. Third, the simulations use independent normally
distributed study estimates with known sampling variances; dependent
effects, rare outcomes, estimated within-study variances, robust
variance estimation, and nonnormal random effects require separate
evaluation. Fourth, the stress experiment covers a wrong-direction
exponential gradient, a U-shaped function, and a step decrease, but it
does not exhaust nonlinear or discontinuous heterogeneity processes.
Fifth, plug-in and mKH intervals still condition on estimated scale
structure and do not fully propagate uncertainty in τ₀² and γ; profile
or bootstrap procedures deserve further finite-sample study. Sixth, the
grid does not establish performance for very small meta-analyses,
extreme sampling-variance imbalance, multivariate outcomes, or complex
dependence structures. Seventh, the empirical illustration has narrow
observed DR support and influential studies, so it demonstrates model
falsifiability and workflow rather than validating a universal
design-quality scale. Eighth, the optimizer audit revealed exact
start-value nonmovement in some low-information conditions; those cells
should be interpreted as weakly identified, and future software should
expose multiple-start or profile diagnostics by default. Finally,
DR-Meta remains a working model for residual scale; it is not a
bias-correction device, and a systematic mean relation must be
represented in the location model.

\subsection{\texorpdfstring{\textbf{11.7 Software
implications}}{11.7 Software implications}}\label{software-implications}

The public drmeta 0.2.2 implementation supports the exact γ = 0
boundary, constrained and unrestricted scale fits, location moderators,
observed-support scale contrasts, classical heterogeneity summaries,
leave-one-out diagnostics, and a parametric-bootstrap utility. The
software implementation should not be read as independent validation of
every inferential procedure: the simulation evidence concerns
estimation, coverage, misspecification, and bound sensitivity rather
than a full finite-sample calibration of the boundary test. Numerical
provenance is therefore kept explicit: Tables 5-6 use the CRAN drmeta
0.2.2 package run, while Tables 1-4/B1 and the empirical reference
calculations use the exact archived profiled-REML reference outputs. The versioned CRAN/GitHub software release and the
manuscript-specific reproducibility deposit serve distinct archival
roles.\textsuperscript{15}

\section{\texorpdfstring{\textbf{12.
Conclusion}}{12. Conclusion}}\label{conclusion}

Design information can matter to the scale as well as the location of
meta-analytic effects, but those roles should not be conflated. DR-Meta
is a constrained location-scale specification for a prespecified
hypothesis that residual heterogeneity does not increase with design
robustness. Its strongest property is not that it always improves
precision; the simulations show that it does not. Its value is that the
directional restriction is explicit, nested at random effects,
mathematically transparent, and directly diagnosable against an
unrestricted alternative. When the restriction is wrong, the constrained
estimator can fall back toward γ = 0. When the direction is
approximately right, even an imperfect variance shape can yield a useful
efficiency gain. The method should therefore be reported with
conventional random effects, unrestricted scale diagnostics, location
models when needed, and interval/bound sensitivity analyses.

\clearpage

\section{\texorpdfstring{\textbf{References}}{References}}\label{references}

1. DerSimonian R, Laird N. Meta-analysis in clinical trials. Control
Clin Trials. 1986;7(3):177-188. doi:10.1016/0197-2456(86)90046-2

2. Hedges LV, Olkin I. Statistical Methods for Meta-Analysis. Academic
Press; 1985.

3. Borenstein M, Hedges LV, Higgins JPT, Rothstein HR. Introduction to
Meta-Analysis. Wiley; 2009.

4. Greenland S, O\textquotesingle Rourke K. On the bias produced by
quality scores in meta-analysis. Biostatistics. 2001;2(4):463-471.
doi:10.1093/biostatistics/2.4.463

5. Herbison P, Hay-Smith J, Gillespie WJ. Adjustment of meta-analyses on
the basis of quality scores should be abandoned. J Clin Epidemiol.
2006;59(12):1249-1256. doi:10.1016/j.jclinepi.2006.03.008

6. Jüni P, Witschi A, Bloch R, Egger M. The hazards of scoring the
quality of clinical trials for meta-analysis. JAMA.
1999;282(11):1054-1060. doi:10.1001/jama.282.11.1054

7. Viechtbauer W, López-López JA. Location-scale models for
meta-analysis. Res Synth Methods. 2022;13(6):697-715.
doi:10.1002/jrsm.1562

8. Blázquez-Rincón D, López-López JA, Viechtbauer W. Performance of
location-scale models in meta-analysis: a simulation study. Behav Res
Methods. 2025;57(4):118. doi:10.3758/s13428-025-02622-5

9. Doi SAR, Barendregt JJ, Khan S. Quality effects model for
meta-analysis. Epidemiol Health. 2015;37:e2015025.
doi:10.4178/epih/e2015025

10. Self SG, Liang KY. Asymptotic properties of maximum likelihood
estimators and likelihood ratio tests under nonstandard conditions. J Am
Stat Assoc. 1987;82(398):605-610. doi:10.1080/01621459.1987.10478472

11. Knapp G, Hartung J. Improved tests for a random effects
meta-regression with a single covariate. Stat Med.
2003;22(17):2693-2710. doi:10.1002/sim.1482

12. Morris TP, White IR, Crowther MJ. Using simulation studies to
evaluate statistical methods. Stat Med. 2019;38(11):2074-2102.
doi:10.1002/sim.8086

13. Kuper P, Miguel C, Cuijpers P, et al. Sample size and geographical
region predict effect heterogeneity in psychotherapy research for
depression: a meta-epidemiological study. J Clin Epidemiol.
2025;183:111779. doi:10.1016/j.jclinepi.2025.111779

14. R Core Team. R: A Language and Environment for Statistical
Computing. R Foundation for Statistical Computing; 2026.

15. Hait S. drmeta: Design-Indexed Location-Scale Meta-Analysis. R
package version 0.2.2. CRAN; 2026. doi:10.32614/CRAN.package.drmeta.
GitHub release v0.2.2.
\url{https://github.com/causalfragility-lab/drmeta/releases/tag/v0.2.2}

16. Austin PC. An introduction to propensity score methods for reducing
the effects of confounding in observational studies. Multivariate Behav
Res. 2011;46(3):399-424. doi:10.1080/00273171.2011.568786

17. Steiner PM, Cook TD, Shadish WR, Clark MH. The importance of
covariate selection in controlling for selection bias in observational
studies. Psychol Methods. 2010;15(3):250-267. doi:10.1037/a0018719

\section{\texorpdfstring{\textbf{Appendix A. Full Mathematical Proofs
and
Derivations}}{Appendix A. Full Mathematical Proofs and Derivations}}\label{appendix-a.-full-mathematical-proofs-and-derivations}

This appendix supplies complete proofs of Propositions 1-3 and a full
derivation of the pseudo-true mean in Equation 7. It also derives the
scale-attenuation contrast used in Section 4.4 and records the boundary
argument underlying Section 4.5. Throughout, let dᵢ = DRᵢ and let ψ =
(τ₀², γ), with vᵢ \textgreater{} 0, τ₀² ≥ 0, and γ ≥ 0. Define the
model-implied total variance and its inverse-variance weight by

\(\sigma ᵢ²(\psi)\  = \ vᵢ\  + \ \tau ₀²\ exp( - \gamma dᵢ),\ \ \ \ \ \ \ \ wᵢ(\psi)\  = \ 1/\sigma ᵢ²(\psi).\)

For a location-model design matrix X of full column rank, let Σ(ψ) =
diag\{σ₁²(ψ), \ldots, σₖ²(ψ)\}. Conditional on ψ, the Gaussian
likelihood is quadratic in β, so the generalized least-squares estimator
is

\(\hat{\beta}(\psi)\  = \ \{ X^{\prime}\Sigma(\psi)\ensuremath{^{-1}}X\} \ensuremath{^{-1}}\ X^{\prime}\Sigma(\psi)\ensuremath{^{-1}}y.\)

In the scale-only model X = 1ₖ, this reduces to the
inverse-total-variance pooled mean

\(\hat{\mu}_{DR}(\psi)\  = \ (\sum ᵢ\ wᵢyᵢ)/(\sum ᵢ\ wᵢ).\)

These identities are used repeatedly below. The proofs concern the model
as stated in the manuscript and do not require an interpretation of DR
as causal.

\subsection{\texorpdfstring{\textbf{A1. Proposition 1: Nesting and
identification}}{A1. Proposition 1: Nesting and identification}}\label{a1.-proposition-1-nesting-and-identification}

\textbf{Statement.} Recall Proposition 1: if γ = 0, DR-Meta is exactly
the conventional random-effects model with τ² = τ₀²; if τ₀² = 0, the
residual between-study variance disappears and the model reduces to the
common-effect model (or fixed-effects meta-regression when X contains
moderators); and if dᵢ = c for every study, the fitted pooled mean is
the conventional random-effects pooled mean with τ²(c) = τ₀² exp(−γc),
while τ₀² and γ are not separately identifiable.

\textbf{Proof.} For independent study estimates, the log likelihood, up
to an additive constant that does not depend on the parameters, can be
written as

\(\ell(\beta,\ \psi)\  = \  - (1/2)\sum ᵢ\lbrack\ log\{\sigma ᵢ²(\psi)\}\  + \ \{ yᵢ\  - \ xᵢ^{\prime}\beta\} ²/\sigma ᵢ²(\psi)\ \rbrack.\)

First set γ = 0. Since exp(0) = 1, for every i we obtain

\(\sigma ᵢ²(\tau ₀²,\ 0)\  = \ vᵢ\  + \ \tau ₀².\)

Therefore Σ(τ₀²,0) = diag(vᵢ + τ₀²), which is precisely the covariance
matrix of the conventional random-effects meta-analysis or
random-effects meta-regression with residual between-study variance τ² =
τ₀². Consequently the Gaussian likelihood, the profiled likelihood in β,
and the REML criterion coincide with their conventional random-effects
counterparts. Conditional on τ₀², the GLS estimator above becomes the
usual inverse-total-variance estimator with weights 1/(vᵢ + τ₀²). Thus γ
= 0 nests the conventional random-effects model exactly, not
approximately.

Second set τ₀² = 0. Because exp(−γdᵢ) is finite and strictly positive
for finite γ and dᵢ,

\(\sigma ᵢ²(0,\ \gamma)\  = \ vᵢ.\)

The between-study random-effect component is then identically zero. If X
= 1ₖ, the estimator is the common-effect inverse-sampling-variance mean.
If X contains prespecified moderators, the estimator is the
corresponding fixed-effects meta-regression GLS estimator. Notice also
that γ disappears from the likelihood when τ₀² = 0, so γ is not
identified on this lower-variance boundary.

Third suppose the design score is constant: dᵢ = c for all i. Define a
single nonnegative scale quantity

\(\eta\  = \ \tau ₀²\ exp( - \gamma c).\)

Then every total variance is

\(\sigma ᵢ²\  = \ vᵢ\  + \ \eta,\)

and hence Σ = diag(vᵢ + η). The likelihood depends on the pair (τ₀², γ)
only through η. In the scale-only case, the conditional pooled estimator
is therefore

\(\hat{\mu}_{DR}\  = \ \{\sum ᵢ\ yᵢ/(vᵢ\  + \ \eta)\}/\{\sum ᵢ\ 1/(vᵢ\  + \ \eta)\},\)

which is exactly the conventional random-effects estimator evaluated at
residual heterogeneity τ² = η = τ²(c). The same statement holds for
meta-regression because the entire covariance matrix is the conventional
random-effects covariance matrix with residual variance η.

It remains to establish nonidentifiability of τ₀² and γ when DR is
constant. Identification requires the map from the parameter pair (τ₀²,
γ) to the distribution of the observed data to be one-to-one. Here that
map factors through the scalar η. If c \textgreater{} 0 and η
\textgreater{} 0, then for every admissible value \ensuremath{\tilde{\gamma}} ≥ 0, the pair

\(\tilde{\tau}_0^2\  = \ \eta\ exp(\tilde{\gamma}c),\ \ \ \ \ \ \ \ \tilde{\gamma}\  \geq \ 0,\)

satisfies \ensuremath{\tilde{\tau}}₀² exp(−\ensuremath{\tilde{\gamma}}c) = η and therefore produces exactly the same Σ and
the same likelihood. There are infinitely many such pairs. If c = 0,
then η = τ₀² and the likelihood is completely independent of γ, so γ is
again unidentified. If η = 0, then τ₀² = 0 and γ is arbitrary, giving
the same boundary model. Thus no constant-score design can separate the
baseline scale τ₀² from the gradient γ. Because REML also depends on the
scale parameters only through Σ, the same nonidentifiability holds for
REML estimation. This proves all parts of Proposition 1. □

\subsection{\texorpdfstring{\textbf{A2. Proposition 2: Conditional
monotonicity of study
weights}}{A2. Proposition 2: Conditional monotonicity of study weights}}\label{a2.-proposition-2-conditional-monotonicity-of-study-weights}

\textbf{Proof.} Suppose two studies i and j have equal sampling
variances vᵢ = vⱼ = v and dᵢ \textgreater{} dⱼ. Their DR-Meta weights
are

\(w(d)\  = \ \lbrack v\  + \ \tau ₀²\ exp( - \gamma d)\rbrack \ensuremath{^{-1}}.\)

Because γ ≥ 0 and dᵢ \textgreater{} dⱼ, multiplication by −γ reverses
the weak order, so −γdᵢ ≤ −γdⱼ. Since the exponential function is
strictly increasing,

\(exp( - \gamma\ d(i))\  \leq \ exp( - \gamma\ d(j)).\)

Multiplying by τ₀² ≥ 0 and adding the common positive sampling variance
v preserves the weak order:

\(v\  + \ \tau ₀²\ exp( - \gamma\ d(i))\  \leq \ v\  + \ \tau ₀²\ exp( - \gamma\ d(j)).\)

Both denominators are strictly positive because v \textgreater{} 0.
Taking reciprocals therefore reverses the inequality, yielding

\(w(i)\  \geq \ w(j).\)

The inequality is strict exactly when the heterogeneity term differs
across the two scores. Given dᵢ \textgreater{} dⱼ, this occurs when γ
\textgreater{} 0 and τ₀² \textgreater{} 0. If γ = 0, both exponentials
equal 1. If τ₀² = 0, the scale term vanishes. In either case wᵢ = wⱼ =
1/v.

The same conclusion follows from differentiation. For d in the score
support,

\(dw(d)/dd\  = \ \gamma\tau ₀²\ exp( - \gamma d)\ /\ \lbrack v\  + \ \tau ₀²\ exp( - \gamma d)\rbrack ²\  \geq \ 0,\)

with strict positivity when γ \textgreater{} 0 and τ₀² \textgreater{} 0.
This derivative establishes monotonicity of the weight function
conditional on a fixed sampling variance.

The equal-v qualification is essential. If vᵢ and vⱼ differ, then wᵢ ≥
wⱼ is equivalent to

\(v(i)\  - \ v(j)\  \leq \ \tau ₀²\{ exp( - \gamma\ d(j))\  - \ exp( - \gamma\ d(i))\}.\)

A sufficiently smaller sampling variance for a lower-DR study can
therefore outweigh the scale advantage associated with higher DR.
Proposition 2 makes only the conditional statement and is therefore
proved. □

\subsection{\texorpdfstring{\textbf{A3. Proposition 3: Fixed-effect
upper bound on raw study
weights}}{A3. Proposition 3: Fixed-effect upper bound on raw study weights}}\label{a3.-proposition-3-fixed-effect-upper-bound-on-raw-study-weights}

\textbf{Proof.} For every study i, τ₀² ≥ 0 and exp(−γdᵢ) \textgreater{}
0. Hence the model-implied residual heterogeneity contribution is
nonnegative:

\(\tau ₀²\ exp( - \gamma dᵢ)\  \geq \ 0.\)

Adding vᵢ \textgreater{} 0 gives

\(vᵢ\  + \ \tau ₀²\ exp( - \gamma dᵢ)\  \geq \ vᵢ\  > \ 0.\)

Because both quantities are positive, taking reciprocals reverses the
first inequality:

\(0\  < \ 1/\lbrack vᵢ\  + \ \tau ₀²\ exp( - \gamma dᵢ)\rbrack\  \leq \ 1/vᵢ.\)

The left-hand side is wᵢ, proving 0 \textless{} wᵢ ≤ 1/vᵢ. If τ₀² = 0,
equality holds for every study. If τ₀² \textgreater{} 0 and γ and dᵢ are
finite, exp(−γdᵢ) is strictly positive, so the denominator is strictly
larger than vᵢ and wᵢ \textless{} 1/vᵢ. The proposition concerns the raw
inverse-total-variance weight. It does not claim that a
study\textquotesingle s normalized percentage weight must be smaller
than its normalized percentage weight under a common-effect analysis,
because normalization changes the denominator across all studies
simultaneously. Proposition 3 is therefore proved. □

\subsection{\texorpdfstring{\textbf{A4. Full derivation of the
pseudo-true constant-mean
target}}{A4. Full derivation of the pseudo-true constant-mean target}}\label{a4.-full-derivation-of-the-pseudo-true-constant-mean-target}

Let D denote DR and V denote the sampling variance. Suppose the true
conditional mean is m(D,V) = E(Y \textbar{} D,V), while the fitted
scale-only model imposes a constant location μ. Hold the
variance-function parameter ψ fixed and define

\(\sigma ²(D,V;\psi)\  = \ V\  + \ \tau ₀²\ exp( - \gamma D),\ \ \ \ \ \ \ \ w(D,V;\psi)\  = \ 1/\sigma ²(D,V;\psi).\)

Assume E{[}wY²{]} \textless{} ∞ and E{[}w{]} \textgreater{} 0,
conditions that allow differentiation under the expectation and
guarantee a finite weighted quadratic criterion. Up to an additive
constant, the population negative Gaussian log likelihood for μ is

\(Q(\mu;\psi)\  = \ (1/2)E\lbrack\ log\{\sigma ²(D,V;\psi)\}\  + \ w(D,V;\psi)(Y\  - \ \mu)²\ \rbrack.\)

The logarithmic term does not depend on μ. Differentiating the quadratic
term yields

\(\partial Q(\mu;\psi)/\partial\mu\  = \ E\lbrack w(D,V;\psi)(\mu\  - \ Y)\rbrack.\)

A second differentiation gives

\(\partial ²Q(\mu;\psi)/\partial\mu ²\  = \ E\lbrack w(D,V;\psi)\rbrack\  > \ 0.\)

Therefore Q is strictly convex in μ and has a unique minimizer. Setting
the first derivative equal to zero gives

\(E\lbrack w(\mu*\  - \ Y)\rbrack\  = \ 0,\)

so

\(\mu \ast \ E\lbrack w\rbrack\  = \ E\lbrack wY\rbrack,\ \ \ \ \ \ \ \ hence\ \ \ \ \ \ \ \ \mu \ast \  = \ E\lbrack wY\rbrack/E\lbrack w\rbrack.\)

Because w is a measurable function of D and V, the law of iterated
expectations gives

\(E\lbrack wY\rbrack\  = \ E\{ w\ m(D,V)\}.\)

Consequently the pseudo-true target is

\(\mu \ast \  = \ E\{ w(D,V;\psi)m(D,V)\}\ /\ E\{ w(D,V;\psi)\}.\)

If the true mean depends only on D, this reduces to the form stated in
Equation 7, μ∗ = E\{w m(D)\}/E{[}w{]}. For a fixed finite collection of
design scores and sampling variances, the corresponding deterministic
weighted target is ∑ᵢ wᵢmᵢ/∑ᵢwᵢ. The derivation shows precisely why
scale reweighting does not generally identify the high-DR mean: μ∗ is a
weighted average of the true conditional means, and the weights depend
jointly on V, τ₀², γ, and D.

The qualification ``for fixed variance parameters'' in Section 3.3 is
important. If ψ is also estimated under misspecification, the joint
pseudo-true parameter (μ∗, ψ∗) minimizes the full expected criterion.
The location component still satisfies the same weighted score equation
evaluated at ψ∗, but ψ∗ must simultaneously satisfy its own population
scale-score equations. Thus Equation 7 is the exact conditional
pseudo-true location target given the variance parameters, not a claim
that the scale parameters remain at their data-generating values under
arbitrary mean misspecification.

For completeness, if a p-dimensional location model X is fitted instead
of a constant mean and E{[}wXX′{]} is nonsingular, the same argument
yields the weighted projection target

\(\beta \ast \  = \ \{ E\lbrack wXX^{\prime}\rbrack\} \ensuremath{^{-1}}\ E\lbrack wX\ m(X,D,V)\rbrack.\)

This is the population GLS projection of the true conditional mean onto
the chosen location model under the design-indexed
inverse-total-variance weighting. □

\subsection{\texorpdfstring{\textbf{A5. Derivation of the
scale-attenuation contrast and score-rescaling
behavior}}{A5. Derivation of the scale-attenuation contrast and score-rescaling behavior}}\label{a5.-derivation-of-the-scale-attenuation-contrast-and-score-rescaling-behavior}

The fitted residual between-study variance at score value d is

\(\tau ²(d)\  = \ \tau ₀²\ exp( - \gamma d).\)

Assume first that τ₀² \textgreater{} 0 and choose dH ≥ dL. The variance
ratio between the higher and lower score values is

\(R\tau(dL,dH)\  = \ \tau ²(dH)/\tau ²(dL)\  = \ exp\{ - \gamma(dH\  - \ dL)\}.\)

The factor τ₀² cancels. Since γ ≥ 0 and dH − dL ≥ 0, the exponent −γ(dH
− dL) is nonpositive. Therefore

\(0\  < \ R\tau(dL,dH)\  \leq \ 1.\)

Defining Aτ = 1 − Rτ gives

\(0\  \leq \ A\tau(dL,dH)\  < \ 1.\)

Equality Rτ = 1 and Aτ = 0 occurs when γ = 0 or dH = dL. If γ
\textgreater{} 0 and dH \textgreater{} dL, then 0 \textless{} Rτ
\textless{} 1 and 0 \textless{} Aτ \textless{} 1. If τ₀² = 0, both
fitted residual variances are zero, so the literal ratio τ²(dH)/τ²(dL)
is 0/0 and has no substantive variance-ratio interpretation; in that
boundary case there is no residual heterogeneity to attenuate.

The contrast also explains why the raw γ coefficient depends on the
numerical scaling of DR. Consider an affine rescaling d′ = a + bd with b
\textgreater{} 0. Since d = (d′ − a)/b,

\(\tau ²(d)\  = \ \tau ₀²\ exp\lbrack - \gamma(d^{\prime}\  - \ a)/b\rbrack\  = \ \{\tau ₀²\ exp(\gamma a/b)\}\ exp\{ - (\gamma/b)d^{\prime}\}.\)

Thus the same variance function can be written in the rescaled score as

\(\tau ²(d^{\prime})\  = \ (\tau_0^{\prime})^2\ exp( - \gamma^{\prime} d^{\prime}),\ \ \ \ \ \ \ \ (\tau_0^{\prime})^2\  = \ \tau ₀²\ exp(\gamma a/b),\ \ \ \ \ \ \ \ \gamma^{\prime}\  = \ \gamma/b.\)

The numerical value of γ changes by the inverse scale factor 1/b, but
the observed-support contrast is invariant because

\(\gamma^{\prime}(dH^{\prime}\  - \ dL^{\prime})\  = \ (\gamma/b)\{ b(dH\  - \ dL)\}\  = \ \gamma(dH\  - \ dL).\)

Hence Rτ = exp\{−γ(dH − dL)\} and Aτ = 1 − Rτ are unchanged by an
equivalent affine re-expression of the score when the model parameters
are transformed consistently. This is why reporting γ together with a
prespecified observed-support variance contrast is more interpretable
than reporting γ alone. □

\subsection{\texorpdfstring{\textbf{A6. Boundary argument for γ =
0}}{A6. Boundary argument for γ = 0}}\label{a6.-boundary-argument-for-ux3b3-0}

The constrained scale parameter satisfies γ ∈ {[}0,∞). When the true
value is γ₀ = 0 and τ₀² \textgreater{} 0, the parameter point lies on
the boundary rather than in the interior of the admissible parameter
space. In particular, no open Euclidean neighborhood of γ₀ is contained
in {[}0,∞): every interval (−ε, ε) contains inadmissible negative
values. The local perturbation set for γ is therefore a half-line rather
than the full real line.

Standard interior likelihood theory obtains a symmetric local quadratic
approximation by allowing local parameter perturbations in all
directions around the true point. That geometric condition fails at γ₀ =
0. Consequently, a two-sided Wald statistic based on an unconstrained
normal approximation and a routine likelihood-ratio comparison using the
ordinary interior chi-square reference distribution are not justified
solely by the standard regular MLE theorem. Boundary likelihood theory
can yield nonstandard mixture limits under additional conditions, but
the exact form must respect the nuisance-parameter structure and any
additional boundary such as τ₀² = 0. The manuscript therefore recommends
boundary-aware resampling rather than asserting a universal
finite-sample reference distribution for the constrained γ test
{[}11{]}.

\section{\texorpdfstring{\textbf{Appendix B. Additional Simulation
Results}}{Appendix B. Additional Simulation Results}}\label{appendix-b.-additional-simulation-results}

\subsection{\texorpdfstring{\textbf{B1. Main factorial results under
lower
heterogeneity}}{B1. Main factorial results under lower heterogeneity}}\label{b1.-main-factorial-results-under-lower-heterogeneity}

When τ₀² = 0.02, differences among RE, constrained DR-Meta, and
unrestricted LS are small because there is comparatively little residual
heterogeneity for a scale model to explain. The table reports the
complete low-heterogeneity grid, including the empirically anchored γ =
1 condition and both plug-in and mKH coverage.

\newpage

Table B1. Main factorial performance under lower heterogeneity (τ₀² =
0.02)

\textbf{Panel A. Point-estimation performance}

\begin{longtable}[]{@{}
  >{\raggedright\arraybackslash}p{(\columnwidth - 8\tabcolsep) * \real{0.1262}}
  >{\raggedright\arraybackslash}p{(\columnwidth - 8\tabcolsep) * \real{0.1650}}
  >{\raggedright\arraybackslash}p{(\columnwidth - 8\tabcolsep) * \real{0.2330}}
  >{\raggedright\arraybackslash}p{(\columnwidth - 8\tabcolsep) * \real{0.2330}}
  >{\raggedright\arraybackslash}p{(\columnwidth - 8\tabcolsep) * \real{0.2427}}@{}}
\toprule\noalign{}
\begin{minipage}[b]{\linewidth}\raggedright
\textbf{γ}
\end{minipage} & \begin{minipage}[b]{\linewidth}\raggedright
\textbf{Fit}
\end{minipage} & \begin{minipage}[b]{\linewidth}\raggedright
\textbf{Bias}
\end{minipage} & \begin{minipage}[b]{\linewidth}\raggedright
\textbf{RMSE}
\end{minipage} & \begin{minipage}[b]{\linewidth}\raggedright
\textbf{RMSE/RE}
\end{minipage} \\
\midrule\noalign{}
\endhead
\bottomrule\noalign{}
\endlastfoot
\multicolumn{5}{@{}>{\raggedright\arraybackslash}p{(\columnwidth - 8\tabcolsep) * \real{1.0000} + 8\tabcolsep}@{}}{%
\textbf{k = 20}} \\
0 & RE & 0.0021 & 0.04407 & 1.000 \\
0 & DR & 0.0022 & 0.04520 & 1.026 \\
0 & LS & 0.0025 & 0.04644 & 1.054 \\
1 & RE & -0.0007 & 0.03851 & 1.000 \\
1 & DR & -0.0009 & 0.03954 & 1.027 \\
1 & LS & -0.0009 & 0.04036 & 1.048 \\
2 & RE & -0.0005 & 0.03547 & 1.000 \\
2 & DR & -0.0006 & 0.03572 & 1.007 \\
2 & LS & -0.0007 & 0.03621 & 1.021 \\
4 & RE & 0.0002 & 0.03022 & 1.000 \\
4 & DR & -0.0001 & 0.03061 & 1.013 \\
4 & LS & -0.0003 & 0.03108 & 1.028 \\
\multicolumn{5}{@{}>{\raggedright\arraybackslash}p{(\columnwidth - 8\tabcolsep) * \real{1.0000} + 8\tabcolsep}@{}}{%
\textbf{k = 50}} \\
0 & RE & -0.0006 & 0.02741 & 1.000 \\
0 & DR & -0.0004 & 0.02779 & 1.014 \\
0 & LS & -0.0005 & 0.02823 & 1.030 \\
1 & RE & -0.0004 & 0.02297 & 1.000 \\
1 & DR & -0.0003 & 0.02301 & 1.002 \\
1 & LS & -0.0003 & 0.02318 & 1.009 \\
2 & RE & -0.0005 & 0.02112 & 1.000 \\
2 & DR & -0.0005 & 0.02128 & 1.007 \\
2 & LS & -0.0006 & 0.02149 & 1.018 \\
4 & RE & 0.0002 & 0.01937 & 1.000 \\
4 & DR & 0.0001 & 0.01926 & 0.994 \\
4 & LS & 0.0001 & 0.01936 & 0.999 \\
\end{longtable}

\newpage

\textbf{Panel B. Interval coverage}

\begin{longtable}[]{@{}
  >{\raggedright\arraybackslash}p{(\columnwidth - 6\tabcolsep) * \real{0.1500}}
  >{\raggedright\arraybackslash}p{(\columnwidth - 6\tabcolsep) * \real{0.1900}}
  >{\raggedright\arraybackslash}p{(\columnwidth - 6\tabcolsep) * \real{0.3300}}
  >{\raggedright\arraybackslash}p{(\columnwidth - 6\tabcolsep) * \real{0.3300}}@{}}
\toprule\noalign{}
\begin{minipage}[b]{\linewidth}\raggedright
\textbf{γ}
\end{minipage} & \begin{minipage}[b]{\linewidth}\raggedright
\textbf{Fit}
\end{minipage} & \begin{minipage}[b]{\linewidth}\raggedright
\textbf{Plug-in coverage}
\end{minipage} & \begin{minipage}[b]{\linewidth}\raggedright
\textbf{mKH coverage}
\end{minipage} \\
\midrule\noalign{}
\endhead
\bottomrule\noalign{}
\endlastfoot
\multicolumn{4}{@{}>{\raggedright\arraybackslash}p{(\columnwidth - 6\tabcolsep) * \real{1.0000} + 6\tabcolsep}@{}}{%
\textbf{k = 20}} \\
0 & RE & 0.930 & 0.933 \\
0 & DR & 0.919 & 0.923 \\
0 & LS & 0.906 & 0.911 \\
1 & RE & 0.949 & 0.950 \\
1 & DR & 0.944 & 0.947 \\
1 & LS & 0.937 & 0.941 \\
2 & RE & 0.942 & 0.951 \\
2 & DR & 0.939 & 0.949 \\
2 & LS & 0.935 & 0.943 \\
4 & RE & 0.963 & 0.963 \\
4 & DR & 0.959 & 0.962 \\
4 & LS & 0.955 & 0.957 \\
\multicolumn{4}{@{}>{\raggedright\arraybackslash}p{(\columnwidth - 6\tabcolsep) * \real{1.0000} + 6\tabcolsep}@{}}{%
\textbf{k = 50}} \\
0 & RE & 0.948 & 0.951 \\
0 & DR & 0.938 & 0.941 \\
0 & LS & 0.932 & 0.935 \\
1 & RE & 0.955 & 0.958 \\
1 & DR & 0.955 & 0.958 \\
1 & LS & 0.955 & 0.957 \\
2 & RE & 0.957 & 0.957 \\
2 & DR & 0.945 & 0.948 \\
2 & LS & 0.940 & 0.945 \\
4 & RE & 0.957 & 0.961 \\
4 & DR & 0.951 & 0.954 \\
4 & LS & 0.946 & 0.950 \\
\end{longtable}

These low-heterogeneity conditions reinforce the identification result:
γ can be unstable or boundary-concentrated while the pooled location
changes little. The practical value of DR-Meta is therefore limited when
residual heterogeneity is already small.

\section{\texorpdfstring{\textbf{Appendix C. Study-Level Data for the
Empirical
Illustration}}{Appendix C. Study-Level Data for the Empirical Illustration}}\label{appendix-c.-study-level-data-for-the-empirical-illustration}

The table reports the aggregate study-level inputs used in Section 8. DR
is defined as Quality/7. The quality criteria are design/process
variables and do not use the realized meta-analytic effect size.

Table C1. Study-level inputs for the 18-study empirical illustration

\begin{longtable}[]{@{}
  >{\raggedright\arraybackslash}p{(\columnwidth - 8\tabcolsep) * \real{0.3603}}
  >{\raggedright\arraybackslash}p{(\columnwidth - 8\tabcolsep) * \real{0.1406}}
  >{\raggedright\arraybackslash}p{(\columnwidth - 8\tabcolsep) * \real{0.1287}}
  >{\raggedright\arraybackslash}p{(\columnwidth - 8\tabcolsep) * \real{0.1853}}
  >{\raggedright\arraybackslash}p{(\columnwidth - 8\tabcolsep) * \real{0.1853}}@{}}
\toprule\noalign{}
\begin{minipage}[b]{\linewidth}\raggedright
\textbf{Study}
\end{minipage} & \begin{minipage}[b]{\linewidth}\raggedright
\textbf{Cohen d}
\end{minipage} & \begin{minipage}[b]{\linewidth}\raggedright
\textbf{SE}
\end{minipage} & \begin{minipage}[b]{\linewidth}\raggedright
\textbf{Quality (0-7)}
\end{minipage} & \begin{minipage}[b]{\linewidth}\raggedright
\textbf{DR = quality/7}
\end{minipage} \\
\midrule\noalign{}
\endhead
\bottomrule\noalign{}
\endlastfoot
Askew et al. & 0.00 & 0.022 & 4 & 0.571 \\
Babcock \& Georgiou & 0.05 & 0.042 & 4 & 0.571 \\
Conway et al. & -1.14 & 0.010 & 5 & 0.714 \\
Cung et al. & 0.22 & 0.032 & 6 & 0.857 \\
Hobbs & -0.36 & 0.032 & 4 & 0.571 \\
Mokher et al. & 0.05 & 0.131 & 5 & 0.714 \\
Ortagus 2018 & 0.27 & 0.113 & 6 & 0.857 \\
Ortagus 2023 & -0.99 & 0.069 & 5 & 0.714 \\
Ortagus et al. 2024 & -0.34 & 0.090 & 6 & 0.857 \\
Paulsen \& McCormick & -0.10 & 0.014 & 5 & 0.714 \\
Richards-Babb et al. & 0.15 & 0.013 & 6 & 0.857 \\
Rogers et al. & 0.02 & 0.123 & 5 & 0.714 \\
Ryan et al. & 0.05 & 0.058 & 5 & 0.714 \\
Shea \& Bidjerano & 0.22 & 0.108 & 4 & 0.571 \\
Sublett & 0.10 & 0.032 & 5 & 0.714 \\
Tadesse & 0.25 & 0.050 & 6 & 0.857 \\
Van Horne et al. & 0.20 & 0.045 & 5 & 0.714 \\
Wladis et al. & -0.78 & 0.326 & 5 & 0.714 \\
\end{longtable}

Note. The smallest supplied standard errors are Conway et al. (SE =
0.010; source evidence-base sample approximately 9,663) and
Richards-Babb et al. (SE = 0.013; source evidence-base sample
approximately 6,114). These large-sample study-level inputs are retained
as coded rather than estimated by DR-Meta. DR is stored computationally
as the full-precision ratio Quality/7 rather than as the three-decimal
display values shown in Table C1. Section 8.2 reports the complete
18-study leave-one-out analysis.

\section{\texorpdfstring{\textbf{Appendix D. Reproducibility
Details}}{Appendix D. Reproducibility Details}}\label{appendix-d.-reproducibility-details}

Computational provenance is experiment-specific. The expanded main
factorial results, mKH comparisons, scale-misspecification experiment,
optimization-bound experiment, and optimizer-start diagnostics use the
exact archived fit-level outputs from an independently implemented
profiled-REML reference run. The score-quality and design-linked
mean-misspecification experiments use R 4.6.0 with drmeta 0.2.2 and
master seed 20260806. The main factorial
uses 1,000 replications per condition; DR values are generated from
Beta(2,2), sampling variances from Uniform(0.005, 0.030), and study
estimates from the normal location-scale model in Section 6.1. The
scale-misspecification experiment uses 1,000 replications per mechanism.
The γ-bound experiment is a separate paired Monte Carlo run: within each
sensitivity replicate the same data are fit with upper bounds 8 and 20,
but those sensitivity datasets are independent of the main factorial
run. RMSE is computed against μ = 0.30. For constrained fits,
near-boundary frequency is \ensuremath{\hat{\gamma}} \textless{} 0.001; for signed unrestricted
fits, a near-zero diagnostic uses \textbar \ensuremath{\hat{\gamma}}\textbar{} \textless{}
0.001.

The empirical illustration uses the same profiled REML objective as the
model derivation. Computational DR values are Quality/7 at full
precision. For the intercept-only model, at fixed τ₀² and γ the location
estimate is the inverse-total-variance weighted mean. The constrained
optimizer permits γ = 0 exactly, while the unrestricted fit uses
symmetric γ bounds to diagnose whether the best-fitting scale relation
has the prespecified sign. The reproducibility bundle includes the exact
Table 7 results and all 18 leave-one-out fits.

The manuscript bundle records the table-by-table engine provenance,
software/version targets, available random-number settings, optimization
bounds, archived raw fit-level results, summary CSV files,
figure-generation code, diagnostic scripts, full-precision empirical
inputs, environment information, and checksums. The archived raw fit-level output is the exact target for the reported
Monte Carlo summaries. Because the original random stream for every
copied stress/bound artifact was not retained during expansion of those
analyses, fresh R reruns of the declared simulation designs are treated
as Monte Carlo replications rather than byte-identical reconstructions. The
bundle is separate from the software archive because the latter
identifies software rather than the exact manuscript analysis.
The manuscript-specific reproducibility bundle is included with this arXiv submission as ancillary material.

\end{document}